\UseRawInputEncoding
\documentclass[10pt,aps,prd,notitlepage,showpacs,nofootinbib,preprintnumbers,superscriptaddress]{revtex4-2}
\usepackage{graphicx}
\usepackage{amsmath}
\usepackage{amssymb}
\usepackage{float}
\usepackage{comment}
\usepackage{slashed}
\usepackage[normalem]{ulem}
\usepackage{braket}
\usepackage{caption}
\usepackage{subcaption}
\usepackage{amsmath}
\usepackage{amssymb}
\usepackage[usenames,dvipsnames]{color}
\usepackage[colorinlistoftodos]{todonotes}
\usepackage[colorlinks=true,citecolor=darkred,urlcolor=darkred, pdfborder={0 0 0}]{hyperref}
\usepackage[normalem]{ulem}
\definecolor{darkred}{rgb}{0.6,0,0}
\usepackage[colorinlistoftodos]{todonotes}
\definecolor{linkcolor}{rgb}{0,0,0.5}
 
\newcommand {\ignore}[1]{}

\def\gsim{\raise0.3ex\hbox{$\;>$\kern-0.75em\raise-1.1ex\hbox{$\sim\;$}}}
\def\lsim{\raise0.3ex\hbox{$\;<$\kern-0.75em\raise-1.1ex\hbox{$\sim\;$}}}

\usepackage{soul}
\definecolor{mightnightblue}{RGB}{25,25,112}
\definecolor{brown}{rgb}{0.59, 0.29, 0.0}

\newcommand{\beq}{\begin{equation}}
\newcommand{\eeq}{\end{equation}}
\newcommand{\bea}{\begin{eqnarray}}
\newcommand{\eea}{\end{eqnarray}}

\def\21{$\mathrm{SU(2)_L \otimes U(1)_Y}$}

\begin{document}
\bibliographystyle{unsrt}   
\title{Charged Higgs induced 5 and 6 lepton signatures from heavy neutrinos at the LHC}
\author{Arindam Das}
\email{adas@particle.sci.hokudai.ac.jp}
\affiliation{Institute for the Advancement of Higher Education, Hokkaido University, Sapporo 060-0817, Japan}
\affiliation{Department of Physics, Hokkaido University, Sapporo 060-0810, Japan}
\author{Shinya Kanemura}
\email{kanemu@het.phys.sci.osaka-u.ac.jp}
\affiliation{Department of Physics, Osaka University
Toyonaka, Osaka 560-0043, JAPAN}
\author{Prasenjit Sanyal}
\email{prasenjit.sanyal@apctp.org}
\affiliation{Asia Pacific Center for Theoretical Physics, Pohang 37673, Republic of Korea}
\date{\today}
\begin{abstract}
We propose an anomaly free gauged U$(1)$ extension of the SM where three right handed heavy neutrinos, being charged under the general U$(1)$ gauge group, are introduced to explain the origin of the tiny neutrino mass through the seesaw mechanism after the general U$(1)$ symmetry is broken. Due to the breaking of the general U$(1)$ symmetry a neutral beyond the standard model gauge boson $Z^\prime$ acquires mass. There are two Higgs doublets in this model where one interacts with the SM fermions and the other one interacts with the right handed heavy neutrinos and charged leptons. The charged multiplet of the second Higgs can completely decay into the heavy neutrinos and charged lepton in the neutrinophilic limit of the model parameters. The charged Higgs pair production can be influenced due to presence of the $Z^\prime$ boson at the High Luminosity LHC (HL-LHC) in addition to the neutral SM gauge bosons. The pair produced charged Higgs bosons decay into SM charged leptons and heavy neutrinos. Following the leading decay modes of the heavy neutrinos into charged leptons and $W$ boson we study the 5 and 6 lepton final states after the leptonic and hadronic decay of the $W$ bosons considering solely muons and electrons in the final state. Combining the electron and muon final states we estimate  the significance of the 5 and 6 charged lepton processes in the $m_{H^\pm}-m_N$ plane for different benchmark points of $m_{Z^\prime}$. It is found that the 5 (6) charged lepton processes could be probed at the High Luminosity LHC (HL-LHC) with at least 5 (3) $\sigma$ significance and there are parameter regions where the significance could be larger. 
\end{abstract}
\maketitle
\preprint{EPHOU-22-015, OU-HET 1151, APCTP Pre2022-019}
\setcounter{page}{1}
\setcounter{footnote}{0}
\section{Introduction}
A strong indication of the existence of beyond the Standard Model (BSM) physics has been indicated by the observation of the neutrino oscillation data \cite{Patrignani:2016xqp} which lead us to extend the SM. There are many proposed ways and amongst them seesaw scenario which could be the simplest proposal where SM is extended by a SM-singlet heavy Majorana Right Handed Neutrinos (RHNs) \cite{Minkowski:1977sc,Mohapatra:1979ia,Schechter:1980gr,Yanagida:1979as,GellMann:1980vs,Glashow:1979nm,Mohapatra:1979ia}. Being SM gauge singlet RHNs mix with the SM light neutrinos to interact with the SM gauge sector. Hence the RHNs being can only be produced at the high energy colliders through the light-heavy mixings which can be naturally small $\mathcal{O}(10^{-6})$ for the TeV scale Majorana type RHNs to reproduce light neutrino mass around 0.1 eV. Attempts have been made to study such TeV scale or lighter Majorana neutrinos  at the Large Hadron Collider (LHC) where same sign dilpeton signature could be distinctively an interesting channel from a variety of production modes \cite{delAguila:2008cj,Mitra:2011qr,Das:2015toa}. With a general parameterization of the Dirac mass matrix \cite{Casas:2001sr} this light-heavy mixing can be larger. However, it becomes less than 0.01 in order to simultaneously satisfy different experimental constraints including the neutrino oscillation data, the electroweak precision measurements and lepton flavor violation \cite{Das:2017nvm}. As a result production of RHNs at the LHC is extremely suppressed. In addition to that RHN can be produced at the electron positron collider in association with neutrinos and in electron proton collider in association with jet. However, those production modes are also suppressed by the square of the mixing. The study of RHN production in these colliders from a variety of final states and the limits on the light heavy mixing has been studied in \cite{Banerjee:2015gca, Antusch:2016vyf, Antusch:2016ejd, Antusch:2017hhu, Chakraborty:2018khw, Das:2018usr, Antusch:2019eiz}. In these studies light-heavy mixing has a strong implication because production cross section is proportional to the mixing squared. Therefore smallness of the light-heavy mixing makes RHN production from the SM gauge bosons become challenging at the high energy colliders. On the other hand RHN neutrinos  can be produced in pair from the neutral current interaction which is suppressed by fourth power of light-heavy mixing. As a result the production cross section further suppressed \cite{Das:2017pvt} making heavy neutrino search more challenging.   

Apart from the seesaw scenario there is another interesting possibility where a gauged B$-$L (Baryon minus Lepton) extension of the SM could be a more compelling scenario \cite{Davidson:1978pm,Marshak:1979fm,Mohapatra:1980qe,Wetterich:1981bx,Masiero:1982fi,Mohapatra:1982xz,Davidson:1987mh,Buchmuller:1991ce,Kanemura:2011vm,Nomura:2019wlo,Nomura:2021adf} to explain the origin of tiny neutrino mass. Here the RHNs play crucial role to cancel the gauge and mixed gauge-gravity anomalies. After the spontaneous breaking of the B$-$L gauge symmetry, the RHNs acquire Majorana masses which automatically implement the seesaw mechanism after the electroweak symmetry breaking. Due to the U$(1)_{\rm{B}-\rm{L}}$ gauge extension there exists a neutral, BSM gauge boson $Z^\prime$ having direct interaction with the RHNs. Due to the nonzero B$-$L charges the $Z^\prime$ boson couples with the SM fermions. As contribution of the seesaw mechanism, the RHNs decay into the SM particles through the light-heavy mixing. There is an alternative approach where SM can be extended by a general U$(1)_X$ gauge group. There are three generations of the RHN present in this scenario in order to solve the mixed gauge and gauge gravity anomalies. As a result the general U$(1)_X$ charges of the left and right handed SM fermions becomes different which is manifested by their interactions with $Z^\prime$ boson showing a chiral nature. A  variety of phenomenological studies have been performed including pair production of heavy neutrino at the high energy colliders from $Z^\prime$ using prompt \cite{Das:2017flq,Das:2017deo}, displaced \cite{Das:2018tbd,Das:2019fee,Chiang:2019ajm} and boosted \cite{Das:2022rbl1} decays of heavy neutrinos, pair production of charged leptons at the electron positron colliders \cite{Das:2021esm} and light $Z^\prime$ searches at beam dump experiments \cite{Asai:2022zxw} respectively. Due to the presence of the general U$(1)_X$ charges the production of the heavy neutrino pair from $Z^\prime$ can be enhanced which has been proposed in \cite{Das:2017flq,Das:2017deo}.

In this paper we consider a general U$(1)_X$ extension of the SM based on SU$(3)_c \otimes$ SU$(2)_L \otimes$ U$(1)_Y \otimes$ U$(1)_X$ where U$(1)_X$ is a generalization of U$(1)_{\rm{B}-\rm{L}}$ such that the U$(1)_X$ charges of the particles are realized as a linear combination of the SM U$(1)_Y$ and U$(1)_{\rm{B}-\rm{L}}$ charges so called non-exotic U$(1)_X$ model \cite{Das:2017flq,Das:2017deo,Appelquist:2002mw}. In the U$(1)_X$ model flavor universal charges are assigned for the three generations of the RHNs \cite{Das:2017flq}, however, there is another charge assignment for the RHNs to make model anomaly free. In this alternative assignment two RHNs have U$(1)_X$ charge $-4$ and the remaining one is assigned with a $+5$ U$(1)_X$ charge. With a minimal extension of the Higgs sector two RHNs with U$(1)_X$ charge $-4$ are involved in a minimal seesaw mechanism while the other RHN with U$(1)_X$ charge $+5$ can not couple with SM particles because of the U$(1)_X$ charges of the SM particles \cite{Das:2017deo}. The connection between additional Higgs doublet and heavy neutrinos have been explored in \cite{Ko:2014tca,Guo:2017ybk,Huitu:2017vye,Tang:2017plx,Abdallah:2021npg,Choudhury:2020cpm,Asai:2020xnz}. Due to the fact of seesaw mechanism light neutrino masses are generated. This scenario also involves a neutral BSM gauge boson $Z^\prime$ which interacts with the RHNs and SM particles of this model and these aspects have been studied in \cite{Das:2017flq,Das:2017deo,Das:2018tbd,Das:2019fee,Chiang:2019ajm,Das:2022rbl1}. 

Apart from $Z^\prime$, there is another interesting fact that the RHNs with U$(1)_X$ charge $-4$, involved in the seesaw mechanism, can interact with the charged multiplet of a second Higgs doublet involved belonged to the extended Higgs sector. We found that dilepton searches give stringent constraints on the U$(1)_X$ gauge coupling. Consider High Luminosity LHC (HL-LHC) at $\sqrt{s}=14$ TeV with 3000 fb$^{-1}$ luminosity we propose a parameter space on charged Higgs and RHN masses at different signal significance.  Due to the fact that all the Higgs fields are charged under the U$(1)_X$ gauge group, they can interact with $Z^\prime$. As a result if kinematically allowed, the charged multiplets of the Higgs bosons can be produced from $Z^\prime$ in pair  and each charged Higgs can further decay in to an RHN and charged lepton. The RHNs can decay into SM gauge bosons and leptons. Depending on the hadronic and leptonic decays of the SM gauge bosons originated after the RHN decay, we obtain multilpeton signature containing five or six leptons in the final state/s in association with missing momentum and jets coming from the SM gauge bosons. Such rare multilepton mode has not been explored in the context of U$(1)_X$ scenario before. Additionally there is another interesting fact in this model due to the presence of general U$(1)_X$ symmetry. The left and right handed fermions differently couple with the $Z^\prime$ due to the presence U$(1)_X$ gauge symmetry. As a result here the $Z^\prime$ has chiral nature and it can be manifested when $Z^\prime$ interacts with quarks and leptons. As a result the charged Higgs pair production from the left handed and right handed quarks will be different. In addition to that the charged Higgs pair production cross section from $Z^\prime$ will depend on the U$(1)_X$ charges of the quarks and the charged Higgs. Due to the dependence on U$(1)_X$ charges, such a chiral behavior can be manifested in charged Higgs production which has not been explored in any previous literature. Depending on the U$(1)_X$ charges, the branching ratio of the $Z^\prime$ into charged Higgs can be maximized comparing with that of $Z^\prime$ into SM charged leptons. The general U$(1)_X$ gauge coupling can be constrained by the dilepton, dijet searches from the LHC \cite{ATLAS:2019erb,CMS:2019tbu,ATLAS:2019bov,CMS:2018mgb}, ATLAS technical design report (TDR) \cite{CERN-LHCC-2017-018}  and LEP-II \cite{ALEPH:2013dgf,Eichten:1983hw,Electroweak:2003ram}. 

The article is arranged as follows. We discuss the model in Sec.~\ref{model}. Signal and backgrounds of the charged Higgs pair production and its decay into multilepton modes through the RHNs at the LHC have been studied in Sec.~\ref{sig}. Finally we conclude the article in Sec.~\ref{conc}. 
\section{Model}
\label{model}
The U$(1)_X$ extension investigated in this article consists of three generations of SM-singlet RHNs with non-universal charge assignments helping to cancel gauge mixed gauge gravity anomalies. First two generations of the RHNs have charge as $-4$ and third generation has U$(1)_X$ charge as $5$. Due to the non-universal charge assignment the scalar sector of the SM needs to be extended with one SU$(2)_L$ doublet $(H_2)$ and three singlet SM-singlet scalers $(\varphi_{A, B, C})$. Depending on the U$(1)_X$ charge assignments of the fermions in the model we find that the RHNs interact only with the extended Higgs sector. The particle content and the corresponding $U(1)_X$ charges of the particles are given as $(x^\prime_f, f= \{ q,~u,~d,~\ell,~e \})$ in Tab.~\ref{tab:charge}. 
\begin{table}[t]
\begin{center}
\begin{tabular}{||c||ccc||rcl|c||c||c||c||c||c||c||}
\hline
\hline
            & SU(3)$_C$ & SU(2)$_L$ & U(1)$_Y$ & \multicolumn{3}{c|}{U(1)$_X$}&$-2$&$-1$&$-0.5$& $0$& $0.5$ & $1$ & $2$  \\
            &&& &&&&U$(1)_{\rm{R}}$& & &B$-$L&&&  \\
\hline
\hline
&&&&&&&&&&&&&\\[-12pt]
&&&&&&&&&&&&&\\
$Q_L^\alpha$    & {\bf 3}   & {\bf 2}& $\frac{1}{6}$ & $x_q^\prime$ 		& = & $\frac{1}{6}x_H^{} + \frac{1}{3}$   &$0$&$\frac{1}{6}$&$\frac{1}{4}$&$\frac{1}{3}$&$\frac{5}{12}$&$\frac{1}{2}$&$\frac{1}{3}$\\
$u_R^\alpha$    & {\bf 3} & {\bf 1}& $\frac{2}{3}$ & $x_u^\prime$ 		& = & $\frac{2}{3}x_H^{} + \frac{1}{3}$   &$-1$&$-\frac{1}{3}$&$0$&$\frac{1}{3}$&$\frac{1}{2}$&$1$&$\frac{5}{3}$\\
$d_R^\alpha$    & {\bf 3} & {\bf 1}& $-\frac{1}{3}$ & $x_d^\prime$ 		& = & $-\frac{1}{3}x_H^{} + \frac{1}{3}$  &$1$&$\frac{2}{3}$&$\frac{1}{2}$&$\frac{1}{3}$&$\frac{1}{6}$&$0$&$-\frac{1}{3}$\\
\hline
\hline
&&&&&&&&&&&&&\\
$L_L^\alpha$    & {\bf 1} & {\bf 2}& $-\frac{1}{2}$ & $x_\ell^\prime$ 	& = & $- \frac{1}{2}x_H^{} - 1$   &$0$&$-\frac{1}{2}$&$-\frac{3}{4}$&$-1$&$\frac{5}{4}$&$-\frac{3}{2}$&$-2$ \\
$e_R^\alpha$   & {\bf 1} & {\bf 1}& $-1$   & $x_e^\prime$ 		& = & $- x_H^{} - 1$   &$1$&$0$&$-\frac{1}{2}$&$-1$&$-\frac{3}{2}$&$-2$&$-3$ \\
\hline
\hline
&&&&&&&&&&&&&\\
$N_{R_{1,2}}$   & {\bf 1} & {\bf 1}& $0$   & $x_\nu^\prime$ 	& = & $-4$  &$-4$&$-4$&$-4$&$-4$&$-4$&$-4$&$-4$ \\
$N_{R_3}$   & {\bf 1} & {\bf 1}& $0$   & $x_\nu^{\prime\prime}$ 	& = & $5$  &$5$&$5$&$5$&$5$&$5$&$5$&$5$ \\
\hline
\hline
&&&&&&&&&&&&&\\
$H_1$         & {\bf 1} & {\bf 2}& $-\frac{1}{2}$  &  $x_{H_{1}}^{}$ 	& = & $-\frac{x_H^{}}{2}$ &$1$&$\frac{1}{2}$&$\frac{1}{4}$&$0$&$-\frac{1}{4}$&$-\frac{1}{2}$&$-1$ \\ 
$H_2$         & {\bf 1} & {\bf 2}& $-\frac{1}{2}$  &  $x_{H_{2}}^{}$ 	& = &  $-\frac{1}{2} x_{H}^{}+3$ &$4$&$\frac{7}{2}$&$\frac{13}{2}$&$3$&$\frac{11}{4}$&$\frac{5}{2}$&$2$ \\ 
$\varphi_A$      & {\bf 1} & {\bf 1}& $0$  &  $ x_{\Phi_{1}}^{}$ 	& = & $+8$ &$+8$&$+8$&$+8$&$+8$&$+8$&$+8$&$+8$  \\ 
$\varphi_B$      & {\bf 1} & {\bf 1}& $0$  &  $x_{\Phi_{2}}^{}$ 	& = & $-10$ &$-10$&$-10$&$-10$&$-10$&$-10$&$-10$&$-10$  \\ 
$\varphi_C$      & {\bf 1} & {\bf 1}& $0$  &  $x_{\Phi_{3}}^{}$ 	& = & $-3$ &$-3$&$-3$&$-3$&$-3$&$-3$&$-3$&$-3$  \\ 
\hline
\hline
\end{tabular}
\end{center}
\caption{
Particle content of the alternative U$(1)_X$ model with general U$(1)_X$ charges before and after anomaly cancellation, and $\alpha = 1,2,3$ stands for three generations of the fermions. The charges of the RHNs are non-universal. Considering different benchmark values of the $x_H^{}$ we obtain different U$(1)_X$ charges of left and right handed fermions of the model manifesting the chiral nature of the model. Here, $x_H^{}=0$ is an alternative B$-$L case, which is a vector-like scenario.}
\label{tab:charge}
\end{table}
The U$(1)_X$ charges of the SM fermions are the same for three generations. The general charges can be related to each other from the following gauge and mixed gauge-gravity anomaly cancellation conditions  
\begin{align}
{\rm U}(1)_X \otimes \left[ {\rm SU}(3)_C \right]^2&\ :&
			2x_q^\prime - x_u^\prime - x_d^\prime &\ =\  0~, \nonumber \\
{\rm U}(1)_X \otimes \left[ {\rm SU}(2)_L \right]^2&\ :&
			3x_q^\prime + x_\ell^\prime &\ =\  0~, \nonumber \\
{\rm U}(1)_X \otimes \left[ {\rm U}(1)_Y \right]^2&\ :&
			x_q^\prime - 8x_u^\prime - 2x_d^\prime + 3x_\ell^\prime - 6x_e^\prime &\ =\  0~, \nonumber \\
\left[ {\rm U}(1)_X \right]^2 \otimes {\rm U}(1)_Y &\ :&
			{x_q^\prime}^2 - {2x_u^\prime}^2 + {x_d^\prime}^2 - {x_\ell^\prime}^2 + {x_e^\prime}^2 &\ =\  0~, \nonumber \\
\left[ {\rm U}(1)_X \right]^3&\ :&
			3({6x_q^\prime}^3 - {3x_u^\prime}^3 - {3x_d^\prime}^3 + {2x_\ell^\prime}^3-{x_e^\prime}^3) - 2 x_\nu^{\prime^3}-x_\nu^{\prime \prime^3}   &\ =\  0~, \nonumber \\
{\rm U}(1)_X \otimes \left[ {\rm grav.} \right]^2&\ :&
			3(6x_q^\prime - 3x_u^\prime - 3x_d^\prime + 2x_\ell^\prime-x_e^\prime)-2 x_\nu^{\prime}-x_\nu^{\prime \prime}  &\ =\  0~, 
\label{anom-f}
\end{align}
respectively. Using SM $\otimes~U(1)_X$ gauge symmetry we write the Yukawa interactions as 
\begin{eqnarray}
- L^{\text{lepton}}_Y &=& \bar{L}_L y_l \tilde{H}_1 e_R + \sum_{i=1}^3 \sum_{j=1}^2 Y_D^{ij} \bar{L}_{Li} H_2 N_{R_j} + \frac{1}{2}\sum_{k=1}^2 Y_{N}^{A,k} \bar{N}^C_{R_{k}}\varphi_A N_{R_k} + \frac{1}{2} Y^B_N \bar{N}^C_{R_3}\varphi_B N_{R_3} + h.c. \nonumber \\
-L^{\text{quark}}_Y &=& \bar{Q}_L y_d \tilde{H}_1d_R + \bar{Q}_L y_u H_1u_R  + h.c.
\label{LYukawa}
\end{eqnarray}
where we find that $N_{R_{1, 2}}$ interact with the new Higgs doublet $(H_2)$ to generate the neutrino Dirac mass term while the SM singlet scalars $(\varphi_{A})$ generate the Majorana mass terms for $N_{R_{1, 2}}$ after the U$(1)_X$ symmetry breaking respectively. The Majorana mass term of $N_{R_3}$ is generated from the VEV of $\varphi_{B}$. Due to the $U(1)_X$ charges no Dirac mass term is generated for $N_{R_3}$ as a result $N_{R_3}$ does not participate in the neutrino mass generation mechanism at the tree level as a result it may be considered as a potential DM candidate. These Yukawa interaction terms impose
\bea
x_{H_1}=-x_{\ell}^\prime+ x_e^{\prime}=-x_{q}^\prime+x_{d}^\prime=x_{q}^\prime-x_{u}^\prime;~~~x_{H_2}=x_{\ell}^\prime-4
\eea
to obtain the U$(1)_X$ charges using the anomaly cancellation given in Eq.~\ref{anom-f}. We assume a basis in which $Y_N^A$ of Eq.~\ref{LYukawa} is diagonal without the loss of generality. 

Apart from the $Z^\prime$ interactions with the fermions of this model, the scalar sector in this model is also interesting. The scalar potential for two Higgs doublets and three SM singlet scalars under the $SU(3)\times SU(2) \times U(1)_Y \times U(1)_X$ gauge group can be given by
\begin{eqnarray}
V &=& m_{H_1}^2 H_1^\dagger H_1 + m_{H_2}^2 H_2^\dagger H_2 + m_{\varphi_A}^2 \varphi^*_A \varphi_A + m_{\varphi_B}^2 \varphi^*_B \varphi_B + m_{\varphi_C}^2 \varphi^*_C \varphi_C + \mu [(H_1^\dagger H_2)\varphi_C + h.c.] \nonumber \\
&+& \lambda_1(H_1^\dagger H_1)^2 + \lambda_2 (H_2^\dagger H_2)^2 + \lambda_3 (\varphi^*_A \varphi_A)^2 +
 \lambda_4 (\varphi^*_B \varphi_B)^2 + \lambda_5 (\varphi^*_C \varphi_C)^2 + \lambda_6 (H_1^\dagger H_1)(H_2^\dagger H_2) \nonumber \\
 &+& \lambda_7(H_1^\dagger H_2) (H_2^\dagger H_1) + \lambda_8 (H_1^\dagger H_1)(\varphi_A^* \varphi_A) + \lambda_9 (H_2^\dagger H_2)(\varphi_A^* \varphi_A) + \lambda_{10} (H_1^\dagger H_1)(\varphi_B^* \varphi_B) \nonumber \\ 
 &+& \lambda_{11} (H_2^\dagger H_2)(\varphi_B^* \varphi_B) + \lambda_{12} (H_1^\dagger H_1)(\varphi_C^* \varphi_C) + \lambda_{13} (H_2^\dagger H_2)(\varphi_C^* \varphi_C) + \lambda_{14} (\varphi_A^* \varphi_A)(\varphi_B^* \varphi_B)\nonumber \\
 &+& \lambda_{15} (\varphi_A^* \varphi_A)(\varphi_C^* \varphi_C) + \lambda_{16} (\varphi_B^* \varphi_B)(\varphi_C^* \varphi_C).
\label{potential}
\end{eqnarray}
We parameterize the scalar fields as 
\begin{eqnarray}
&& H_1 =\left(
\begin{array}{c}
\frac{v_1 + \varphi_3 +i\varphi_4}{\sqrt{2}} \\
\varphi^-_1\\
\end{array}
\right), \hspace{0.25cm}  
H_2 =\left(
\begin{array}{c}
\frac{v_2 + \varphi_7 + i\varphi_8}{\sqrt{2}}\\ \hspace{0.25cm}  
\varphi_2^-\\
\end{array} 
\right),~ 
\varphi_i = \frac{v_i + \varphi_{i_R} + + i \varphi_{i_I}}{\sqrt{2}},
\label{scalars}
\end{eqnarray}
with $j=A, B, C,$ where $v_{1,2}$ and $v_{\rm (A, B, C)}$ are the VEVs of the scalar fields in the model satisfying the condition $v_1^2+v_2^2=(246 \rm GeV)^2$. 
For simplicity we consider the mixed quartic couplings between the scalars to be very small so that the Higgs doublet sector can be effectively separated from the singlet Higgs sector allowing the higher order mixing effect between the RHNs after the $U(1)_X$ symmetry breaking to be highly suppressed. Hence the singlet and doublet scalars communicate through the $\mu [(H_1^\dagger H_2)\varphi_C + h.c.]$. Using this term we further impose 
\bea
x_{H_2}=x_{H_1}+3 
\eea 
on the U$(1)_X$ charges of Tab.~\ref{tab:charge} to obtain the final form. The charge assignment of the particles after the anomaly cancellation is given in Tab.~\ref{tab:charge} where the charge assignment for the SM fermions are linear combination of the U$(1)_Y$ and U$(1)_{\rm{B}-\rm{L}}$. We find that for changing $x_H=-2,~-1,~-0.5,~0.5,~1, ~2$ the left and right handed charges under the U$(1)_X$ differ which is observed in their interactions with $Z^\prime$ manifesting its chiral nature. In case of $x_H=-2$ the left handed fermions do not interact with $Z^\prime$ which is an alternative U$(1)_R$ scenario. For $x_H=-0.5$ and 1 right handed up and down type quarks do not interact with $Z^\prime$. Finally for $x_H=-1$, the right handed electron do not interact with $Z^\prime$.  For $x_H=0$ we notice that the fermion sector and $H_1$ in the model manifest the B$-$L charge assignment. Therefore we call it an alternative B$-$L scenario. We consider the collider constraints $v_A^2+v_B^2+v_C^2 \gg v_1^2+v_2^2$ the triple coupling has no significant effect on determining $v_1$, $v_2$, $v_c$ when we arrange the parameters in the scalar potential to have $v_A \sim v_B \sim v_C$ and $\mu < v_A$. When $\varphi_C$ generates VEV, then mixing mass term between $H_1$ and $H_2$ are generated. Hence the potential of the Higgs doublet sector effectively becomes the potential for the Higgs potential of the two Higgs doublet model. Since there is no mixing mixing mass term among $\varphi_{A, B, C}$, there exists two physical Nambu-Goldstone (NG) modes in our model which are not physically dangerous. We fix the SM singlet scalars to be heavier than $Z^\prime$ boson so that $Z^\prime$ boson can not decay into the NG modes.
\subsection{Scalar Sector}
The vacuum expectation values (VEVs) can be obtained by minimizing the scalar potential given by $\partial V/\partial v_1 = \partial V/\partial v_2 = \partial V/\partial v_j = 0$ where $j=A,B,C$.
Minimizing the scalar potential given in Eq.~\ref{potential} using the scalar field from Eq.~\ref{scalars} we obtain the following stationary conditions 
\begin{eqnarray}
&& m_{H_1}^2 + v_1^2 \lambda_1 + \frac{1}{2}v_C^2 \lambda_{12} + \frac{1}{2}v_2^2 \lambda_6 + \frac{1}{2}v_2^2 \lambda_7 + \frac{1}{2}v_A^2 \lambda_8 + \frac{1}{9}v_B^2 \lambda_{10} + \frac{v_2 v_C \mu}{\sqrt{2}v_1}=0 \nonumber \\
&& m_{H_2}^2 + v_2^2 \lambda_2 + \frac{1}{2}v_B^2 \lambda_{11} + \frac{1}{2} v_C^2 \lambda_{13} + \frac{1}{2}v_1^2 \lambda_6 + \frac{1}{2}v_1^2 \lambda_7 + \frac{1}{2}v_A^2 \lambda_9 + \frac{v_1 v_C \mu}{\sqrt{2}v_2}=0
\nonumber \\
&& m_{\varphi_A}^2 + v_A^2 \lambda_3 + \frac{1}{2} v_B^2 \lambda_{14} + \frac{1}{2}v_C^2 \lambda_{15} + \frac{1}{2} v_1^2 \lambda_8 + \frac{1}{2}v_2^2 \lambda_9 =0  \nonumber \\
&& m_{\varphi_B}^2 + v_B^2 \lambda_4 + \frac{1}{2}v_2^2 \lambda_{11} + \frac{1}{2}v_A^2 \lambda_{14} + \frac{1}{2} v_C^2 \lambda_{16} + \frac{1}{2}v_1^2 \lambda_{10} = 0 \nonumber \\
&& m_{\varphi_C}^2 + v_C^2 \lambda_{5} + \frac{1}{2}v_1^2 \lambda_{12} + \frac{1}{2} v_2^2 \lambda_{13} + \frac{1}{2} v_A^2 \lambda_{15} + \frac{1}{2}v_B^2 \lambda_{16} + \frac{v_1 v_2 \mu}{\sqrt{2}v_C}=0 
\end{eqnarray}
Using the minimization conditions, the mass matrix for the charged scalars is given by 
\begin{eqnarray}
L\supset -\Big( \frac{\lambda_7}{2} + \frac{1}{\sqrt{2}}\frac{v_C \mu}{v_1 v_2}\Big)
\left(
\begin{array}{c}
\varphi_1^+ \\
\varphi_2^+ \\
\end{array}
\right)^T
\left(
\begin{array}{cc}
v_2^2 & -v_1 v_2\\
-v_1 v_2 & v_1^2\\
\end{array}
\right)
\left(
\begin{array}{c}
\varphi_1^- \\
\varphi_2^- \\
\end{array}
\right),
\end{eqnarray}
which when diagonalized to obtain the mass of the charged Higgs  $(H^\pm)$ and massless charged Goldstone boson $G^\pm$ such that
\begin{eqnarray}
m_{H^\pm}^2 = -\Big( \frac{\lambda_7}{2} + \frac{1}{\sqrt{2}}\frac{v_C \mu}{v_1 v_2}\Big)v^2,
\end{eqnarray} 
and we transform the charged counterparts in the following way 
\begin{eqnarray}
\left(
\begin{array}{c}
G^\pm \\
H^\pm \\
\end{array}
\right)
=\left(\begin{array}{cc}
\cos\beta & \sin\beta \\
-\sin\beta & \cos\beta \\
\end{array}
\right)
\left(
\begin{array}{c}
\varphi_1^{\pm} \\
\varphi_2^{\pm} 
\end{array}
\right),
\end{eqnarray}
where $\tan\beta=v_2/v_1$ and $v^2=v_1^2 + v_2^2$. The mass matrix of the CP-odd scalar sector is a 3$\times$3 symmetric matrix which can be diagonalized to obtain two neutral Goldstone bosons $G_{1,2}$ which are eaten up by $Z,~Z^\prime$ gauge bosons and $A$ as the CP-odd neutral scalar. The CP-even scalar sector has five physical degrees of freedom $(\varphi_{3,7,A_R, B_R, C_R})$ with a $5\times5$ symmetric matrix which can be diagonalized numerically to obtain five CP even scalars $\chi_{1,2,3,4}$ and $h$. We identify $h$ as the lightest observed Higgs boson of mass 125 GeV \cite{CMS:2012qbp,ATLAS:2012yve}. The mass splitting between charged Higgs and other neutral scalars can lead to interesting scenarios for example the bosonic decay modes of charged like $H^{\pm} \to \chi_{1,2,3,4} (h, A) W^+$ can overcome the neutrinophilic behavior of charged Higgs. Several studies of bosonic decays of charged Higgs are done in two Higgs doublet model \cite{Coleppa:2014cca,Kling:2015uba,Enberg:2018pye,Wang:2021pxc,Mondal:2021bxa,Kim:2022nmm,Kim:2022hvh}. However, we chose the simplest scenario where these bosonic decays are suppressed either kinematically or by suitable mixing angles so that the neutrinophilic nature of charged Higgs gets prominence. We refrain further study on the scalar sector assuming only the neutrinophilic scenario of charged Higgs.
In this context we write down the constraints from the stability of the scalar potential as \cite{Ivanov:2018jmz,Bian:2017xzg}
\begin{eqnarray}
\lambda_{1,2,5} > 0, ~~~ 2\sqrt{\lambda_1 \lambda_2} + \lambda_6 > 0, ~~~ 2\sqrt{\lambda_1 \lambda_2} + \lambda_6 + \lambda_7 > 0, \nonumber \\
2\sqrt{\lambda_1\lambda_5} + \lambda_{12} > 0, ~~~ 2\sqrt{\lambda_2\lambda_5} + \lambda_{13} > 0, \nonumber \\
\sqrt{(\lambda_{12}^2 - 4 \lambda_1 \lambda_5)(\lambda_{13}^2 - 4 \lambda_2 \lambda_5)} + 2\lambda_5 \lambda_6 > \lambda_{12}\lambda_{13}, \nonumber \\    
\sqrt{(\lambda_{12}^2 - 4 \lambda_1 \lambda_5)(\lambda_{13}^2 - 4 \lambda_2 \lambda_5)} + 2\lambda_5 (\lambda_6 + \lambda_7) > \lambda_{12}\lambda_{13}.
\end{eqnarray}
and the conditions for perturbativity and unitarity as \cite{Kanemura:1993hm,Kanemura:2015ska,Bian:2017xzg}
\begin{eqnarray}
|\lambda_{1,2,5}| \leq 4\pi, ~~~ |\lambda_6| \leq 8\pi, ~~~ |\lambda_{12,13}| \leq 8\pi, \nonumber \\
|\lambda_6 \pm \lambda_7| \leq 8\pi, ~~~ |\lambda_6 + 2\lambda_7| \leq 8\pi, \nonumber \\
\Big|\sqrt{\lambda_6(\lambda_6 + 2\lambda_7)}\Big| \leq 8\pi, ~~~ \Big| \lambda_1 + \lambda_2 \pm \sqrt{(\lambda_1 - \lambda_2)^2 + \lambda_7^2} \Big| \leq 8\pi, \nonumber \\
a_{1,2,3} \leq 8\pi
\end{eqnarray}
where $a_{1,2,3}$ are solutions of the cubic equation 
\begin{eqnarray}
x^3 - 2x^2 (3\lambda_1 + 3 \lambda_2 + 2 \lambda_5) - x(2\lambda_{12}^2 + 2\lambda_{13}^2 - 36\lambda_1 \lambda_2 - 24\lambda_1 \lambda_5 - 24\lambda_2\lambda_5 + 4 \lambda_6^2 + 4 \lambda_6 \lambda_7 + \lambda_7^2) \nonumber \\
+ 4 (3\lambda_2 \lambda_{12}^2 - \lambda_{13}(\lambda_7 + 2\lambda_6) + 3\lambda_1\lambda_{13}^2 + \lambda_{5}((\lambda_7 + 2\lambda_6)^2 - 36 \lambda_1 \lambda_2))  = 0.
\end{eqnarray}
\subsection{Gauge Sector}
The kinetic terms of the scalar fields are 
\begin{eqnarray}
L_{kin}= (D_\mu H_1)^\dagger (D^\mu H_1) + (D_\mu H_2)^\dagger (D^\mu H_2) + (D_\mu \varphi_A)^\dagger (D^\mu \varphi_A)
 &+& (D_\mu \varphi_B)^\dagger (D^\mu \varphi_B) + (D_\mu \varphi_C)^\dagger (D^\mu \varphi_C), 
\end{eqnarray}
where the covariant derivatives of the scalar fields can be written as
\begin{eqnarray}
D_\mu H_1 &=& \Big( \partial_\mu + i g \frac{\tau^a}{2} W_{\mu}^a - \frac{1}{2} ig' B_\mu + i g^{\prime\prime}X_{H_1} B'_{\mu} \Big) H_1, \nonumber\\
D_\mu H_2 &=& \Big( \partial_\mu + i g \frac{\tau^a}{2} W_{\mu}^a - \frac{1}{2} ig' B_\mu + i g^{\prime\prime}X_{H_2} B'_{\mu} \Big) H_2, \nonumber \\
D_\mu \varphi_A &=& \partial_\mu + i g^{\prime\prime} X_{\varphi_A}B'_\mu, \nonumber \\
D_\mu \varphi_B &=& \partial_\mu + i g^{\prime\prime} X_{\varphi_B}B'_\mu, \nonumber \\
D_\mu \varphi_C &=& \partial_\mu + i g^{\prime\prime} X_{\varphi_C}B'_\mu,
\label{cov}
\end{eqnarray}
and $X_{H_1},~X_{H_2},X_{\varphi_A},X_{\varphi_B},~X_{\varphi_C}$ are the $U(1)_X$ charges of the the fields $H_1,~H_2,~\varphi_A,~\varphi_B,\varphi_C$ respectively as given in Table~\ref{tab:charge}. Here $g,~g'$ and $g^{\prime\prime}$ are the gauge couplings associated with the $SU(2)_L,~U(1)_Y$ and $U(1)_X$ gauge symmetries respectively. The mass of $W^\pm$ gauge bosons are given by $m_{W^\pm}^2 =\frac{1}{4} g^2 v^2$ and the mass matrix for the neutral gauge boson is given by 
\begin{eqnarray}
L_{\text{gauge}}^{\text{mass}} = \frac{1}{8}
\left(\begin{array}{c}
W_\mu^3 \\
B_\mu \\
B^\prime_\mu \\
\end{array}\right)^T
\left(\begin{array}{ccc}
g^2 v^2 & -gg' v^2 & 2g g^{\prime\prime} a \\
-gg' v^2 & g^{\prime 2}v^2 & -2g' g^{\prime\prime} a \\
2g g^{\prime\prime}a & -2g' g^{\prime\prime}a & 4 g^{\prime\prime 2} b
\end{array}\right) 
\left(\begin{array}{c}
W_\mu^3 \\
B_\mu \\
B^\prime_\mu \\
\end{array}\right),
\end{eqnarray}
where we defined the parameters $a=v_1^2 X_{H_1}^2 + v_2^2 X_{H_2}^2$ and $b=v_1^2 X_{H_1}^2 + v_2^2 X_{H_2}^2 + v_A^2 X_{\varphi_A}^2 + v_B^2 X_{\varphi_B}^2 + v_C^2 X_{\varphi_C}^2$ respectively. Rotating $(W_\mu^3,~B_\mu)^T$ by Weinberg angle to obtain the massless photon $A_\mu$ such that
\begin{eqnarray}
\left(\begin{array}{c}
W^3_\mu \\
B_\mu \\
\end{array}\right) =
\left(\begin{array}{cc}
\cos\theta_w & \sin\theta_w \\
-\sin\theta_w & \cos\theta_w\\
\end{array}\right)
\left(\begin{array}{c}
\tilde{Z}_\mu \\
A_\mu \\
\end{array}\right),
\end{eqnarray}
where $\cos\theta_W = g/\sqrt{g^2 + g'^2}$ and $\sin\theta_W = g'/\sqrt{g^2 + g'^2}$ as in SM. We obtain the mass term of the neutral gauge bosons in the basis of $(\tilde{Z}_\mu,~B'_\mu)$ basis as
\begin{eqnarray}
L_{\text{gauge}}^{\text{mass}} &=& \frac{1}{2}
\left(\begin{array}{c}
\tilde{Z}_\mu \\
B^\prime_\mu \\
\end{array}\right)^T
\left(\begin{array}{cc}
M_{Z,SM}^2 & \Delta^2 \\
\Delta^2 & M_{Z'}^2
\end{array}\right)
\left(\begin{array}{c}
\tilde{Z}_\mu \\
B^\prime_\mu \\
\end{array}\right), \nonumber \\
\Delta^2 &=& \frac{1}{2}g^{\prime\prime}a\sqrt{g^2 + g^{\prime 2}},~~ M_{Z^\prime}^2 = g^{\prime\prime 2}b
\end{eqnarray} 
The mass matrix can be diagonalized by rotation matrix to obtain the neutral gauge bosons $(Z_\mu, ~Z'_\mu)$
\begin{eqnarray}
 \left(\begin{array}{c}
\tilde{Z}_\mu\\
B^\prime\mu\\
\end{array}\right) &=&
\left(\begin{array}{cc}
\cos\theta & \sin\theta\\
-\sin\theta & \cos\theta \\
\end{array}\right)
\left(\begin{array}{c}
Z_\mu\\
Z^\prime\mu\\
\end{array}\right), \nonumber \\
\tan 2\theta &=& \frac{2 \Delta^2}{M_{Z'}^2 - M_{Z,SM}^2}.
\end{eqnarray}
The physical masses of the neutral gauge bosons are 
\begin{eqnarray}
m_Z^2 &=& \frac{1}{2}\Big[M_{Z,SM}^2 + M_Z^{\prime 2} - \sqrt{(M_{Z'}^2 - M_{Z,SM}^2)^2 + 4\Delta^4} \Big], \nonumber \\
m_{Z^{\prime}}^2 &=& \frac{1}{2}\Big[M_{Z,SM}^2 + M_Z^{\prime 2} + \sqrt{(M_{Z'}^2 - M_{Z,SM}^2)^2 + 4\Delta^4} \Big].
\end{eqnarray}
In the limit $\Delta^2 \to 0$ we approximate $m_Z \approx M_{Z,SM}$ and $m_{Z'} \approx M_{Z'}$. Here we have ignored the kinetic mixing of $U(1)_Y$ and $U(1)_X$ gauge bosons assuming that the mixing effect is small \cite{Carena:2004xs}.
\subsection{Neutrino Mass}
\begin{figure}[t!]
\includegraphics[width=160mm]{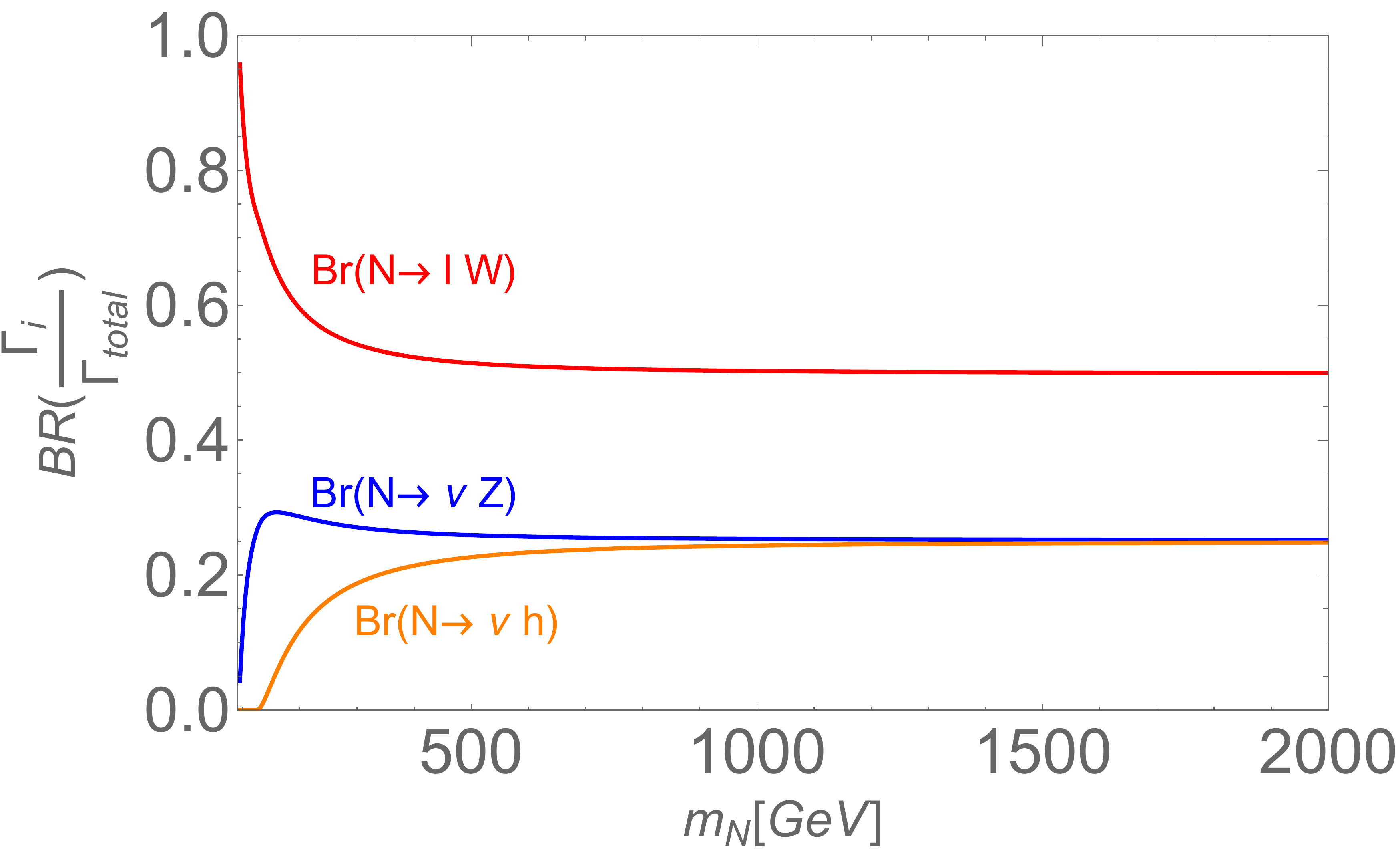}
\caption{Branching ratio of the heavy neutrinos.}
\label{BR-N}
\end{figure}

The mass term for the neutral fermions coming from the Yukawa Lagrangian in Eq.~\ref{LYukawa} after the $U(1)_X$ symmetry is spontaneously broken. Hence the neutrino mass term can be written as 
\begin{eqnarray}
-L_Y^{\text{mass}}= \frac{1}{2}\left(
\begin{array}{c}
\bar{\nu}_L \\ \bar{N}_R^C 
\end{array}\right)^T
\left(\begin{array}{cc}
0 & m_D \\
m_D^T & m_N\\
\end{array}\right)
\left(
\begin{array}{c}
\nu^C_L \\ N_R 
\end{array}\right)
+ \frac{1}{2} m_B \bar{N}_{R_3}^C N_{R_3} + h.c.,
\end{eqnarray} 
where 
\begin{eqnarray}
m_D^{ij} = \frac{Y_D^{ij}v_2}{\sqrt{2}},~m_A^k = \frac{Y^{A,k}_N v_A}{\sqrt{2}},~
m_B = \frac{Y_N^B v_B}{\sqrt{2}}, 
\end{eqnarray}
where the $m_D^{ij}$ is the Dirac mass term generated after the electroweak symmetry breaking and $m_N$ is the Majorana mass term generated after the $U(1)_X$ symmetry breaking. Finally we write the neutrino mass matrix as 
\begin{equation}
\mathcal{M} = \left(\begin{array}{cc}
0 & m_D \\
m_D^T & m_N\\
\end{array}\right),
\label{neutrino}
\end{equation} 
where the first two generations of RHNs, $N_{R_{1,2}}$ are involved whereas the third generation, which can be treated as a potential dark matter candidate, gets the mass $m_B$. 
Assuming the seesaw hierarchy $m_N >> m_D$ and diagonalizing Eq.~\ref{neutrino} to the light neutrino mass as $m_\nu \simeq -m_D m_N^{-1}m_D^T $ where we define $\mathcal{R}= m_D m_N^{-1}$ as the mixing between the light and heavy neutrinos. Due to the light-heavy neutrino mixing, a light neutrino flavor eigenstate ($\nu_\alpha$) is a linear combination of the light ($\nu_m$) and heavy ($N_m$) neutrino mass eigenstates 
\bea 
  \nu_\alpha \ \simeq \ \mathcal{U}_{\alpha m} \nu_m  + \mathcal{R}_{\alpha i} N_i \, ,  
\eea 
where $\mathcal{U}$ is the PMNS mixing matrix to leading order (if we ignore the non-unitarity effects for simplicity). Hence the charged-current (CC) interaction in the lepton sector is then given by  
\bea 
\mathcal{L}_{\rm CC} \ = \ 
 -\frac{g}{\sqrt{2}} W_{\mu}
  \bar{\ell} \gamma^{\mu} P_L 
   \left[ \mathcal{U}_{\alpha m} \nu_m+  \mathcal{R}_{\alpha i} N_i \right] + {\rm H. c.}, 
\label{CC}
\eea
where $g$ is the $SU(2)_L$ gauge coupling and  $P_L =(1- \gamma_5)/2$ is the left-chiral projection operator. Similarly, the neutral-current (NC) interaction is given by 
\bea 
\mathcal{L}_{\rm NC} = -\frac{g}{2 \cos\theta_w}  Z_{\mu} \left[ (\mathcal{U}^\dag \mathcal{U})_{m n} \bar{\nu}_m \gamma^{\mu} P_L \nu_n + (\mathcal{U}^\dag \mathcal{R})_{mi} \bar{\nu}_m\gamma^\mu P_L N_i + (\mathcal{R}^\dag \mathcal{R})_{mi}\bar{N}_m\gamma^\mu P_L N_i\right] + {\rm H. c.} , 
\label{NC}
\eea
where $\theta_w$ is the weak mixing angle. In this analysis we consider a  neutrinophilic scenario. The small neutrino mass originated from the seesaw mechanism requires the Dirac Yukawa coupling to be small which further requires $\tan\beta \leq 10^{-3}$ to reproduce the neutrino oscillation data. When we consider the RHNs are heavier than the SM bosons so that they can decay into $\ell W$, $\nu_{\ell} Z$, and $\nu_{\ell} h$ on-shell modes. The corresponding partial decay widths are given by
\begin{eqnarray}
\Gamma(N_i \rightarrow \ell_{\alpha} W)
 &=& \frac{|\mathcal{R}_{\alpha i}|^{2}}{16 \pi} 
\frac{ (m_{N_i}^2 - M_W^2)^2 (m_{N_i}^2+2 M_W^2)}{m_{N_i}^3 v_h^2} ,
\nonumber \\
\Gamma(N_i \rightarrow \nu_\alpha Z)
&=& \frac{|\mathcal{R}_{\alpha i}|^{2}}{32 \pi} 
\frac{ (m_{N_i}^2 - M_Z^2)^2 (m_{N_i}^2+2 M_Z^2)}{m_{N_i}^3 v_h^2} ,
\nonumber \\
\Gamma(N_i \rightarrow \nu_\alpha h)
 &=& \frac{|\mathcal{R}_{\alpha i}|^{2}}{32 \pi}\frac{(m_{N_i}^2-M_h^2)^2}{m_{N_i} v_h^2},
\label{eq:dwofshell}
\end{eqnarray}
respectively where $\mathcal{R}_{\alpha i}$ is the corresponding light-heavy mixing element. The corresponding branching ratios are shown in Fig.~\ref{BR-N}.
\section{Signal and background analysis}
\label{sig}
In this model the charged Higgs $(H^\pm)$ couples with the SM gauge bosons, $Z^\prime$ through the scalar covariant derivatives given in Eq.~\ref{cov} which involve the corresponding $U(1)_Y$ and $U(1)_X$ charges. At the LHC we consider pair production of $H^\pm$ from the neutral gauge bosons. Due to the neutrinophilic nature of $H_2$, the charged Higgs $H^\pm$ decays into the $\ell^\pm N$ mode. Considering the decays of the $N$ we obtain $5$ and $6$ charged lepton final states in association with jets and/or missing momentum.  
\subsection{Constraints on the model parameters}
In the context of our model scenario the charged Higgs interaction with the quark sector can be given by
\begin{eqnarray}
L \supset -\frac{\sqrt{2}}{v}\bar{u}\Big[V_{CKM} m_d P_R  - m_u V_{CKM}P_L  \Big]d H^+ \tan\beta + h.c. .
\end{eqnarray}
The interaction is similar to Type-I two Higgs doublet model with $\cot\beta$ replaced with $\tan\beta$. The low values of $\tan\beta$, i.e., $v_1 < v_2$ are more favorable to satisfy the seesaw hierarchy by explaining the small neutrino mass. As a result the charged Higgs comes dominantly from the second Higgs doublet and the quark sector Yukawa couplings with charged Higgs are highly suppressed. In LHC the searches of charged Higgs are motivated by QCD induced $H^\pm tb$ associated production. Thereafter $H^\pm$ decays into fermionic or bosonic channels. The suppressed coupling/s of $H^\pm$ to the quark sector reduces the strong production of charged Higgs. Hence there are no exclusion limits on $H^\pm$ parameter space. With similar arguments we can evade the flavor physics constraints like $b\to s \gamma$ \cite{Sanyal:2019xcp}.

\begin{figure}[t!]
\includegraphics[width=85mm]{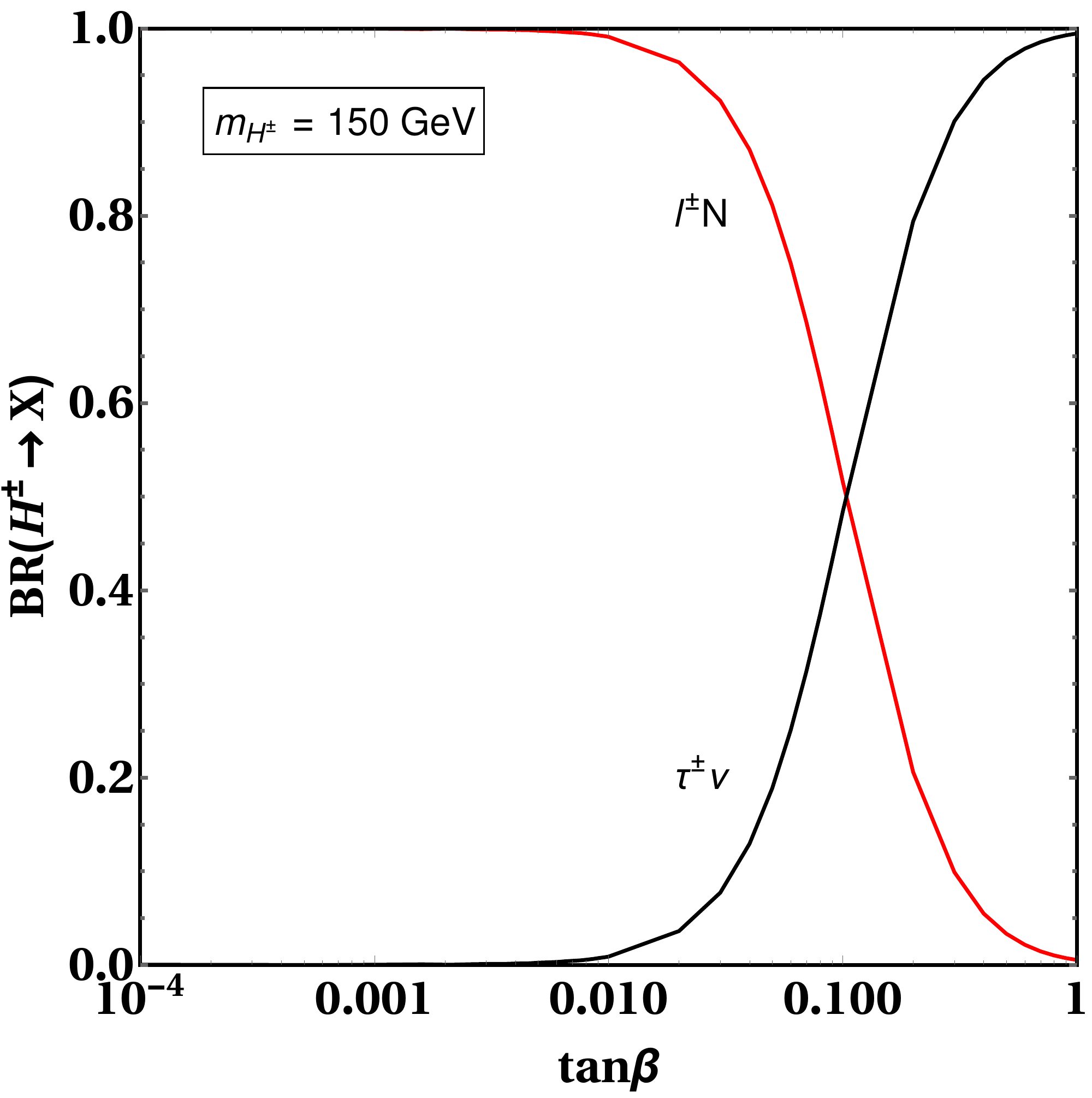}
\includegraphics[width=85mm]{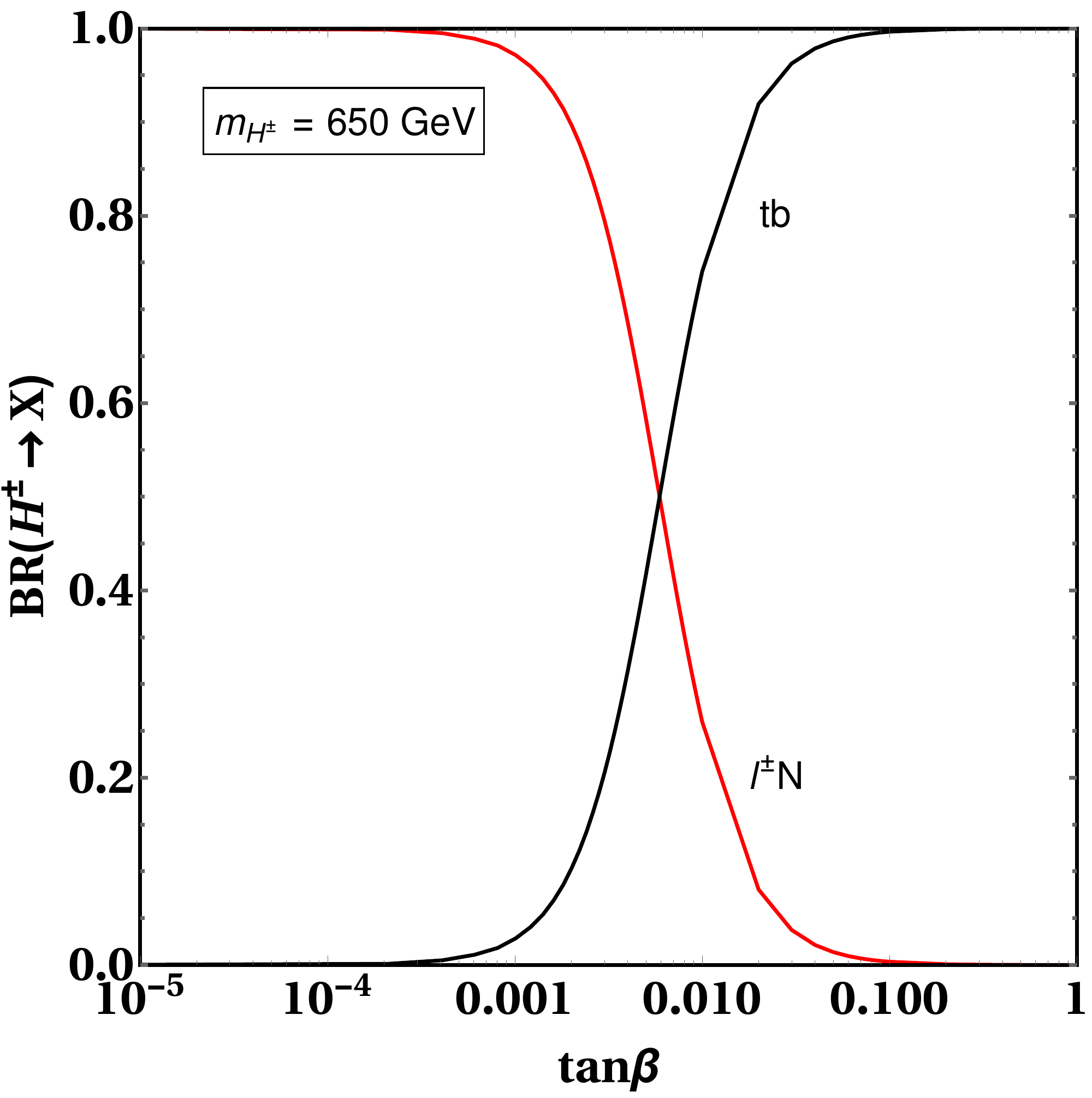}
\caption{Branching ratios of the charged Higgs into different modes as a function of $\tan\beta$ with $m_{N_{1,2}}=100~(300)$ GeV in the left (right) panel.} 
\label{ch-H-1}
\end{figure}

For charged Higgs heavier than top quark, the fermionic decay mode of the charged Higgs manifests a dominant decay into $H^\pm \to tb$ and $H^\pm \to \ell N_i$ channels. Other than the fermionic channel, charged Higgs can decay into the bosonic channels $W^\pm h,\chi_{1,2,3,4},A$. The mass degeneracy between the $H^\pm$, $\chi_{1,2,3,4}$ and $A$ due to the $T$-parameter constraint restricts the decay $H^\pm\to W^\pm\chi_{1,2,3,4},A$. By suitable choice of the mixing angles $\alpha_{1,2,3}$ we completely turn off $H^\pm\to W^\pm h$ decay mode. In our analysis we consider charged Higgs decay only into the fermionic sector. From Eq.~\ref{LYukawa}, we get the charged Higgs interaction with leptons and right handed neutrinos as
\begin{eqnarray}
L \supset -\sum_{i=1}^3 y_\ell^{ii}\sin\beta \bar{\nu}_{iL}H^+ \ell_{iR} + \sum_{i=1}^3\sum_{j=1}^2 Y_D^{ij}\bar{\ell}_{Li} \cos\beta H^- N_{R_j} + h.c. .
\label{eq}
\end{eqnarray} 
Depending on the choice of the $\tan\beta$ and $m_{H^\pm}$ charged Higgs decays into $tb$ mode or $\tau \nu$ mode along with the $\ell^\pm N$ signature. For the low $\tan \beta$ the second term of Eq.~\ref{eq} dominates over the other decay modes since it is proportional to $\cos\beta$. Hence charged Higgs leptonic interaction is dominant into leptons and heavy neutrinos ($N_{1,2}$). The charged Higgs-quark coupling is proportional to $\tan\beta$ hence $H^\pm$ becomes neutrinophilic for $\tan\beta<<1$. The partial decay width of the charged Higgs into the $tb$ channel is given as
\begin{eqnarray}
\Gamma(H^+ \rightarrow t \bar{b})&=&N_C \frac{G_F m_{H^\pm}|V_{tb}|^2}{4\sqrt{2}\pi} \lambda\Big( \frac{m_t^2}{m^2_{H^\pm}},\frac{m_b^2}{m^2_{H^\pm}}\Big)^{1/2}\Big\lbrace(m_t^2 + m_b^2)  \nonumber \\ 
&\times& \Big(1-\frac{m_t^2 + m_b^2}{m^2_{H^\pm}}\Big) + \frac{4m_t^2 m_b^2}{m^2_{H^\pm}}  \Big\rbrace \tan^2\beta, \nonumber \\
\Gamma(H^+ \rightarrow \tau^+ \nu)&=& \frac{G_F m_{H^\pm}{m_\tau}^2}{4\sqrt{2}\pi}\sin^2\beta\Big(1-\frac{m^2_\tau}{m^2_{H^\pm}}\Big)^2,
 \end{eqnarray}
where $N_C(=3)$ is the color factor, $m_t$ is the top quark mass, $\lambda(x,y)^{1/2} = (1+ x^2 + y^2 -2x -2y -2xy)^{1/2}$ is the kinematic factor, $G_F$ is the Fermi constant and $V_{tb}$ is the CKM matrix element respectively. For the neutrinophilic region the decay width of charged Higgs into charged lepton and heavy neutrino is  
\begin{eqnarray}
\Gamma(H^+ \to \ell_{i}^+ N_{j}) = \frac{Y_D^{{ij}^2} \cos^2\beta m_H^{\pm}}{64\pi}\Big(1- \frac{m_{N_j}^2}{m_{H^\pm}^2}\Big)^2,
\end{eqnarray}
where $i=1,2,3$ and $j=1,2$. In Fig.~\ref{ch-H-1} we show the branching ratio comparison between these two channels for different benchmark points.  We find that when $m_{H^\pm}$ is less than top quark mass $(m_{t})$ (left panel), then $m_{H^\pm}$ can decay into $\ell^\pm N$ and $\tau^\pm \nu$ modes where $\ell^\pm N$ mode is dominant when $\tan\beta < 0.1$. Similar behavior is found in the case where  $m_{H^\pm}$ is greater than $m_{t}$ (right panel) where dominant production of the $\ell^\pm N$ mode can be obtained for $\tan\beta < 0.007$. Hence the charged Higgs becomes neutrinophilic below this limit. Hence in the further analysis we consider $\tan\beta= 10^{-4}$. In the neutrinophilic region $H^\pm$ will completely decay into the $\ell^\pm N$ mode. 

\begin{figure}[t!]
\includegraphics[width=85mm]{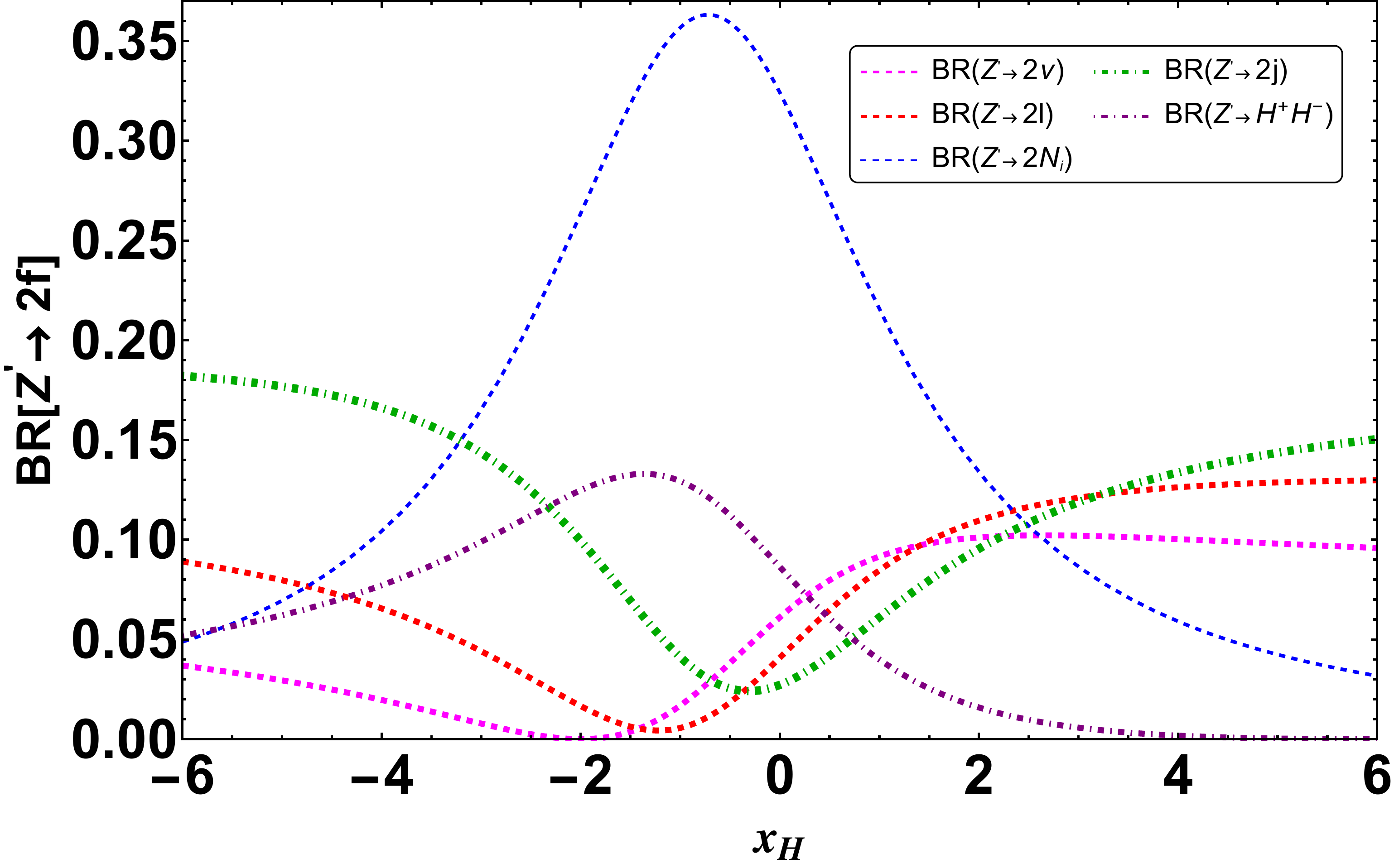}
\includegraphics[width=90mm]{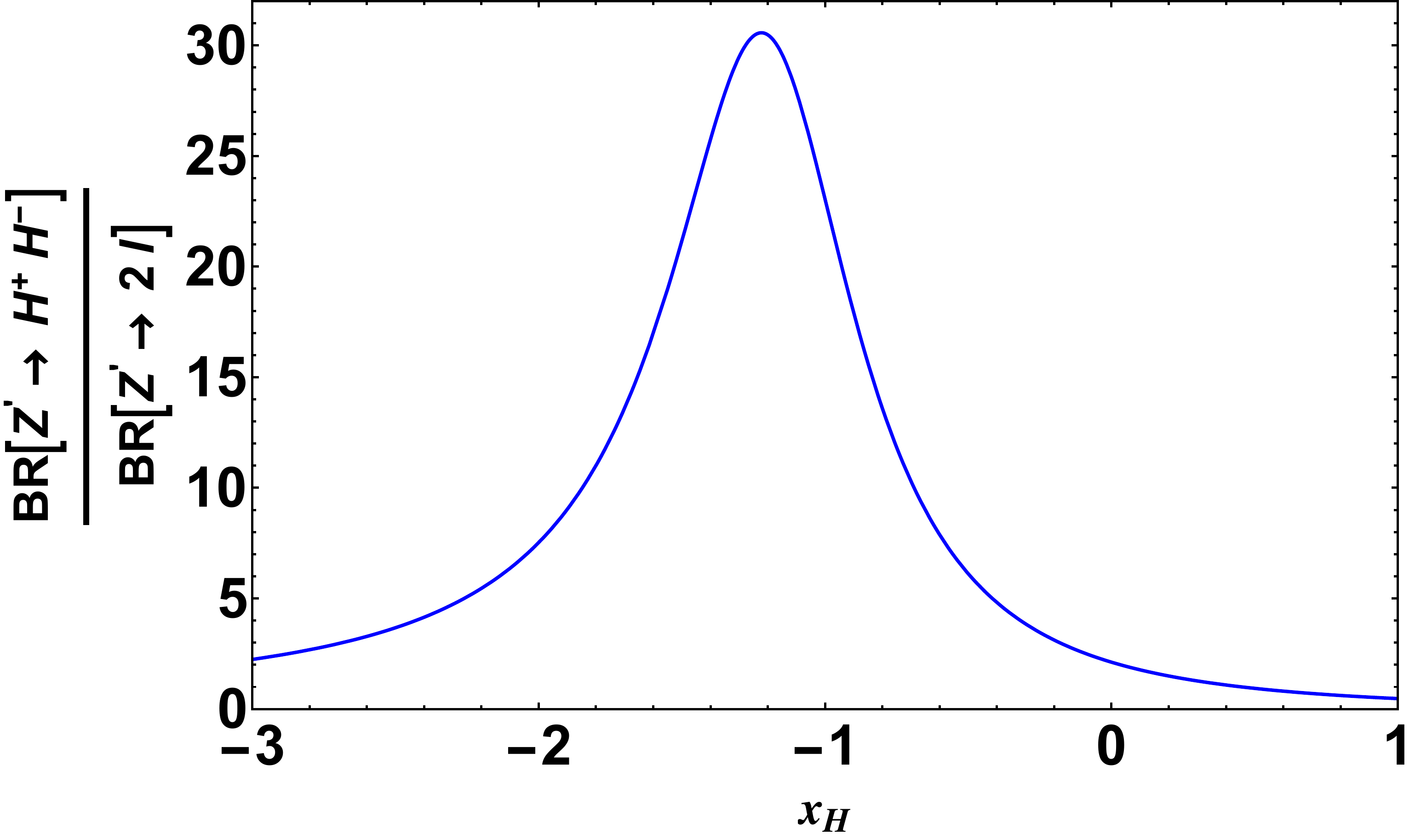}
\caption{ Branching ratio of $Z'$ into different modes (left). The ratio of BR$(Z^\prime \to H^+ H^-)$ to BR$(Z^\prime \to \ell^+ \ell^-)$ considering $M_{Z^\prime}=3$ TeV, $m_{H^\pm}=200$ GeV, $M_{N_{1,2}}=100$ GeV, $\tan\beta=10^{-4}$ (right). }
\label{fig:model-1}
\end{figure}

The partial decay width of $Z' \to H^+ H^-$ in the limit $m_{Z'} >> m_Z$ and $\sin\theta \to 0$ is given by
\begin{eqnarray}
\Gamma(Z' \to H^+ H^-) = \frac{m_{Z'}}{48\pi}\Big[ g^{\prime\prime}(x_{H_1} \sin^2\beta + x_{H_2} \cos^2\beta) \Big]^2 \Big(1 - \frac{4m^2_{H^\pm}}{m^2_{Z'}}\Big)^{3/2}.
\end{eqnarray}
The partial decay widths of the $Z^\prime$ into a pair of charged fermions and heavy neutrinos can be given by
\begin{eqnarray}
\Gamma(Z^\prime \to \overline{f^{L(R)}} f^{L(R)})=  N_c \frac{{g^{\prime\prime}}^2}{24 \pi} {Q_{f}^{L(R)}}^2 m_{Z^\prime},
\label{dec1}
\end{eqnarray}
and
\begin{eqnarray}
\Gamma(Z^\prime \to N_i N_i)&=& \frac{{g^{\prime \prime}}^2}{24 \pi} {Q_N}^2 M_{Z^\prime} \Big(1-4\frac{m_{N_i}^2}{m_{Z^\prime}^2}\Big)^{\frac{3}{2}},
\label{dec2}
\end{eqnarray}
respectively. The corresponding branching ratios of the $Z^\prime$ are shown in the left panel of Fig.~\ref{fig:model-1} considering negligible $Z-Z^\prime$ mixing taking \cite{Carena:2004xs} into account. We find that $Z^\prime$ can dominantly decay into a pair of heavy neutrinos and the next dominant decay mode can be considered as $Z^\prime \to H^+ H^-$ respectively. 
In the right panel we find that BR$(Z^\prime \to H^+ H^-)$ is dominant over the dilepton decay mode from $Z^\prime$ at $x_H=-1.24$ which is considered to study the charged Higgs pair production. As a result if we consider $x_H=-1.24$, the $Z^\prime$ couplings with the left and right handed SM fermions will be different showing the chiral nature of $Z^\prime$. In addition to that branching ratio of $Z^\prime$ into charged Higgs pair will be 30 times more than that of the dilepton mode. As a result the dominant constraints from the dilepton search at the LHC will be less constrained for $x_H=-1.24$.  
Fixing $x_H=-1.24$ we find the U$(1)_X$ charge of the quark doublet is 0.54, that for the right handed up and down type quarks are 1.16 and -0.08 respectively whereas the U$(1)_X$ charges for the left handed and right handed leptons are $-1.16$ and $-2.24$ respectively. The U$(1)_X$ charge of $H_1$ and $H_2$ are $-0.62$ and $2.38$ respectively. Hence the chiral nature of the model persists for the charge assignment when charged Higgs pair production dominates over the dilepton mode.

\begin{figure}[t!]
\includegraphics[width=160mm]{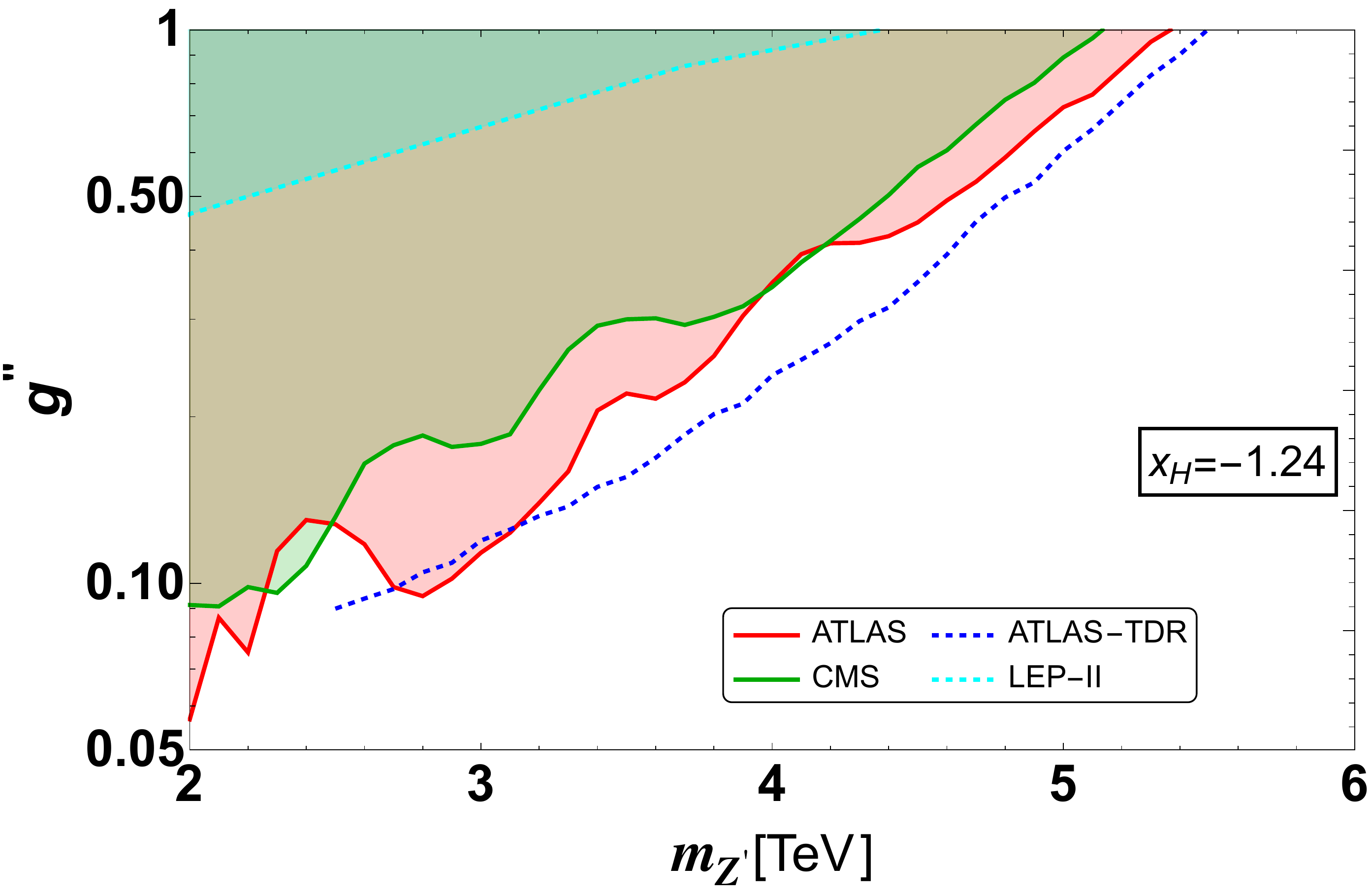}
\caption{Limits on the U$(1)_X$ coupling as a function of $m_{Z^\prime}$ where the shaded region is ruled by the existing experimental data.}
\label{fig:allowed gpp-1}
\end{figure}

To estimate these bounds from the LHC we have considered the case when all the RHNs are heavier than $\frac{M_{Z^\prime}}{2}$ which are shown by the dashed lines by solid Red, solid Green and dashed Blue to represented the LHC dilepton searches from ATLAS-dilepton, CMS-dilepton and ATLAS-TDR (2-electron) respectively. Due to this fact the $Z^\prime$ can not decay into the RHNs which will enhance the other decay modes of $Z^\prime$ allowing the strongest bound on $g^{\prime \prime}$ from the dilepton. We use narrow width approximation at 139(140) $\rm fb^{-1}$ luminosity at the LHC where the $Z^\prime$ production cross section is proportional to ${g^{\prime \prime}}^2$. Hence we estimate the bounds on $g^{\prime \prime}$ using 
\begin{eqnarray}
g^{\prime \prime} = g^{\prime \prime}_{\rm Model} \sqrt{\frac{\sigma_{\rm (ATLAS/ CMS)}}{\sigma_{\rm Model}}},
\end{eqnarray}
where $g^{\prime \prime}_{\rm Model}$ is the coupling considered to estimate the dilepton production cross section from our model. At the High Luminosity LHC (HL-LHC) of 3000 fb$^{-1}$ we can scale the limits from ATLAS (CMS) considering $M_{N} > \frac{M_{Z^\prime}}{2}$ case and obtain that the limit may get uniformly 0.215 (0.216) times stronger than the current luminosity following 
\begin{eqnarray}
g^{\prime \prime} \simeq g^{\prime \prime}_{\rm current} \sqrt{\frac{139(140)~{\rm fb}^{-1}}{\mathcal{L_{\rm future}~{\rm fb^{-1}}}}},
\end{eqnarray}
at a future luminosity of $\mathcal{L_{\rm future}}$. The limits are shown in Fig.~\ref{fig:allowed gpp-1}.
In the similar way we estimate the projected limits on the $g^{\prime \prime}-m_{Z^\prime}$ plane from the ATLAS technical design report (TDR) \cite{CERN-LHCC-2017-018} and the LEP-II limits for $x_H=-1.24$ are estimated following \cite{Das:2021esm}.

\begin{figure}[t!]
\includegraphics[width=125mm]{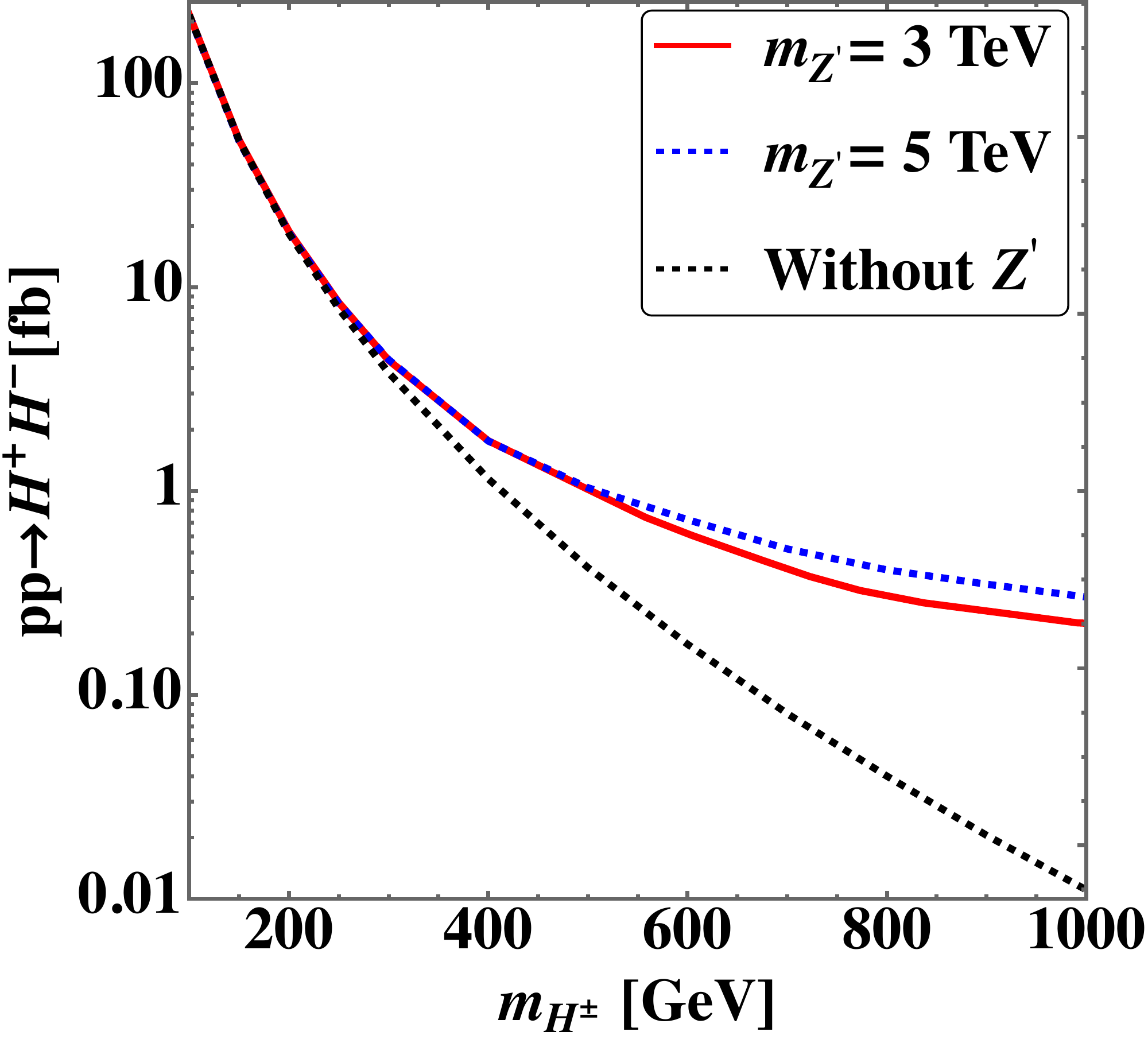}
\caption{Leading order $H^\pm$ production cross section at the 14 TeV LHC as a function of $m_{H^\pm}$ for $m_{Z^\prime}=3$ TeV and 5 TeV respectively considering $x_H=-1.24$, $\tan\beta=10^{-4}$ and $m_N=1$ TeV. The cross section without the influence of $Z^\prime$ falls rapidly at high $m_{H^\pm}$.}
\label{ch-H-production}
\end{figure}

We show the production cross sections of the charged Higgs pair production at the 14 TeV LHC in our model from the $s-$ channel photon, $Z$ and $Z^\prime$ mediated processes including the interference effect. In Fig.~\ref{ch-H-production} we show the charged Higgs pair production at leading order as a function of $m_{H^\pm}$ for $m_{Z^\prime}=3$ TeV and 5 TeV considering $g^{\prime \prime}=0.114$ and $0.615$ respectively from Fig.~\ref{fig:allowed gpp-1} using $x_H=-1.24$, $\tan\beta=10^{-4}$ and $m_{N_{1,2}}=1$ TeV. In this analysis we consider the $3\times 2$ Dirac Yukawa coupling matrix as $Y_{D}^{ii}=0.01$ and $Y_{D}^{ij} \leq 10^{-4}$ where the first (second) entry stands for $i=j~(i\neq j)$. The third generation of the heavy neutrinos do not participate in the neutrino mass generation mechanism and it can be a potential DM candidate. The cross section of the charged Higgs pair production for different $m_{Z^\prime}$ changes from $m_{H^\pm} \geq 500$ GeV. Due to larger $g^{\prime \prime}$ the production cross section is larger in case of $m_{Z^\prime}=5$ TeV than 3 TeV along with the effects of photon and $Z$ mediated processes.  As a reference we present the cross section without $Z^\prime$ induced processes by black dashed line which falls rapidly with large charged Higgs masses.
\subsection{Signal Background Analyses}
In this paper we consider the charged Higgs pair production and the charged Higgs dominantly decays into $\ell^\pm N$ mode. We consider each heavy neutrino decays into dominant mode $N\to \ell W$ as depicted in Fig.~\ref{BR-N} followed by the hadronic decay of one of the $W$ bosons and leptonic decay of the other. Hence we find a final state of 5 charged lepton with two jets in association with missing momentum. On the other hand there is another possibility where both the $W$ bosons decay into leptons giving rise to a completely leptonic final state with 6 charged leptons in association with missing momentum. Hence we concentrate only on the leading decay mode of the heavy neutrinos. Implementing the model in FeynRules \cite{Christensen:2008py, Alloul:2013bka} we generate the events using MadGraph \cite{Alwall:2011uj, Alwall:2014hca} applying CTEQ6L\cite{Pumplin:2002vw} parton distribution function fixing the factorization scale $\mu_F$ as the default MadGraph option followed by the showering, fragmentation and hadronization of the signal and SM backgrounds by the PYTHIA8\cite{Sjostrand:2007gs}. Finally the detector simulation of the showered events were performed by Delphes \cite{deFavereau:2013fsa}. 
\subsubsection{Five charged lepton final state}
\begin{figure}[h]
\includegraphics[width=89mm]{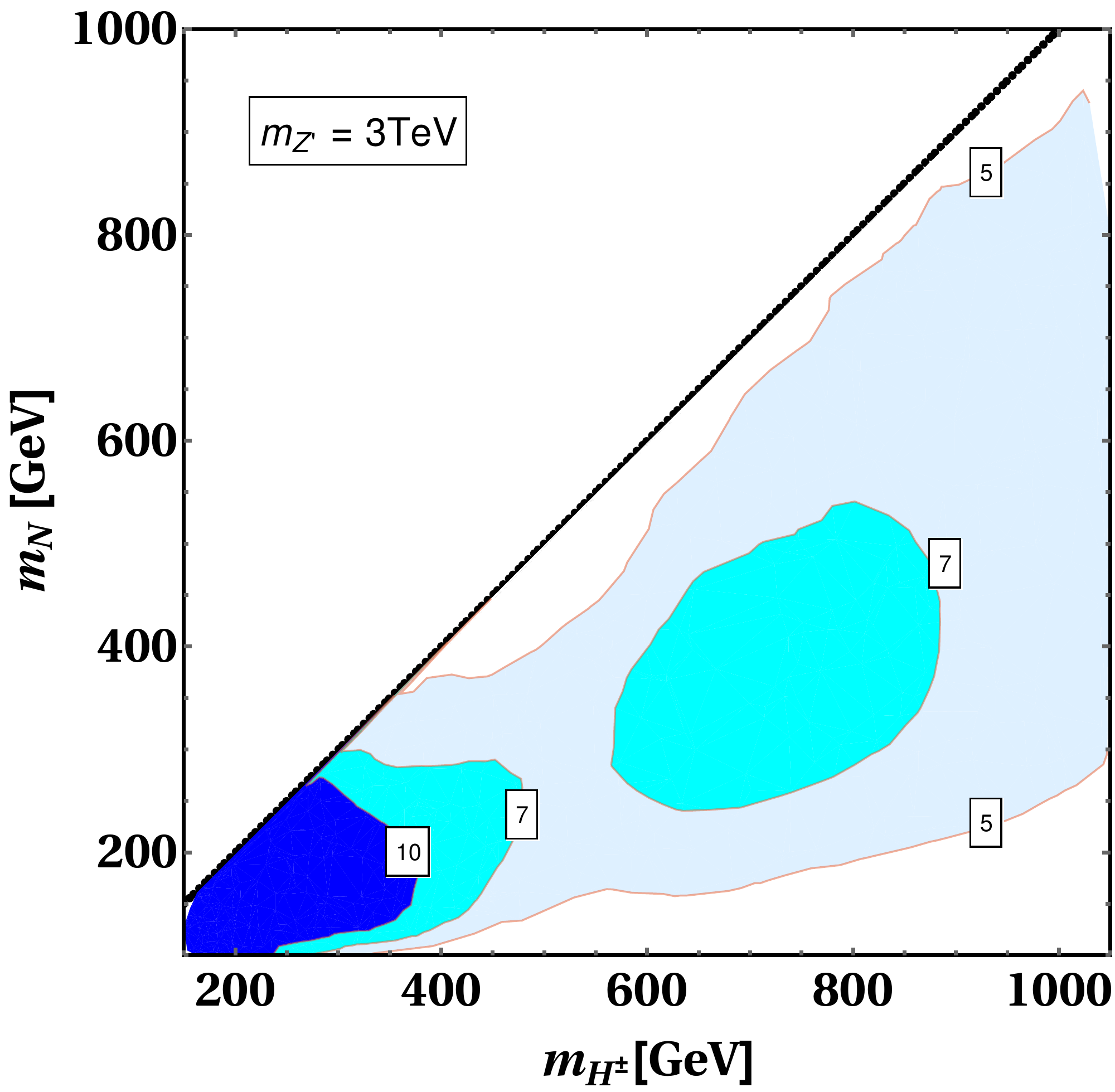}
\includegraphics[width=89mm]{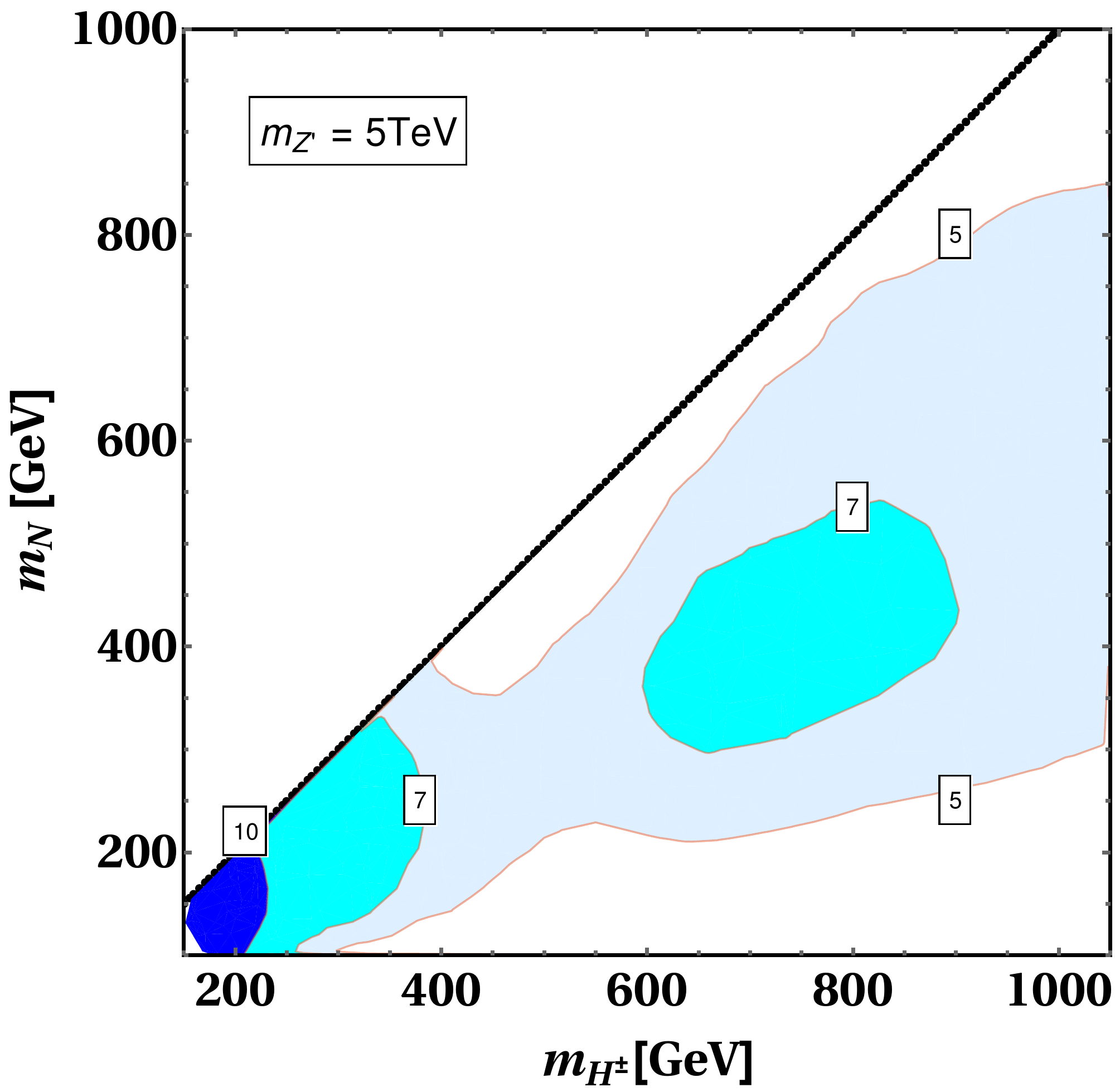}
\caption{Different significance contours for the 5 charged lepton final state with 3000 fb$^{-1}$ luminosity. The black straight line represents $m_{H^{\pm}}=m_N$, below which the on-shell decay of $H^\pm$ occurs.}
\label{fig:allowed gpp-1-1}
\end{figure}

First we discuss the 5 charged lepton final state from the RHN pair production induced by $H^\pm$. We generate irreducible backgrounds for 5 charged leptons which could be fully muons or fully electrons.
In this case we have the events with 5 charged leptons, two jets in association with missing momentum. The irreducible backgrounds are: (i) $ZZW^{\pm}$ where $Z$ decays leptonically and $W^\pm$ decays leptonically, (ii) $t\tilde{t}ZZ$ where $Z$ decays leptonically, $t\tilde{t}$ produce $W^\pm$ with b-jets and one $W$ decays leptonically and the remaining one decays hadronically, (iii) $t\tilde{t}WZ$: the $W$ and $Z$ decay leptonically and $t\tilde{t}$ produce leptonic decay from the $W^\pm$ boson with b-jets, (iv) $ZZW^+W^-$: the $Z$ bosons decay leptonically followed by the leptonic decay of one $W$ boson whereas the other $W$ boson decays hadronically and (v) $ZW^\pm W^\pm W^\mp$: the $Z$ and $W$ bosons decay leptonically to produce a five lepton final state in association with missing momentum. The jets in this events come from the initial state radiations. In addition to the irreducible backgrounds we mention about a reducible background in the form of $ZZ+$jets. In this case four charged leptons are coming from the $Z$ bosons and other lepton can be misidentified from the jets. In this case a $Z$-veto could be applied to reduce this background. In our analysis we did not include $N\to \nu Z$ mode therefore we safely ignore this background. 
To study the signal and backgrounds at the 14 TeV LHC we impose the following selection cuts:
\begin{enumerate}
\item {\it Basic Cuts:} We select events with 5 leptons (fully muons/electrons) in the final state with minimum transverse momentum $p_T(\ell) >$ 10 GeV and $|\eta_\ell|<$ 2.4. Minimum transverse momentum for the accompanying radiated jets to be, $p_T(j)> $ 20 GeV and $|\eta_j|<$ 2.5.
\item {\it Momentum cuts:} To reduce the backgrounds further we impose a hard cut on the transverse momentum of the leptons. We impose $p_T(\ell_{1,2}) > 30$ GeV and $p_T(\ell_{3,4,5}) > 20$ GeV respectively where leptons are ordered according to the transverse momentum. Along with that we impose a missing momentum cut of $p_T^{\rm miss} > 30$ GeV to reduce events with jets as fake leptons further.
\item {\it Lepton separation cuts:} We impose the lepton-lepton separation in the $\eta-\phi$ plane as $\Delta R_{\ell \ell} > 0.4$ and if there are any accompanied jets, then $\Delta R_{\ell j} > 0.4$ and $\Delta R_{j j} > 0.4$ have been considered where $\Delta R = \sqrt{\Delta \eta^2 + \Delta \phi^2}$.   
\end{enumerate}
Additional cuts like $b$-veto and $Z$-veto reduce the top quark and $Z$ boson mediated backgrounds significantly, however since the backgrounds are already negligible due to the low cross sections, the above mentioned cuts are sufficient enough. The background cross sections based on the above cuts are mentioned in Tab.~\ref{BG:5muons} for the case of 5 fully muons/electrons final states. For the signal we choose the benchmark parameters as $m_{Z^\prime}=3$ TeV and 5 TeV, $m_{H^\pm}=$ 150 GeV and 650 GeV with $m_N=100$ GeV and 300 GeV respectively.  We recognize Benchmark Points (BPs): BP1 as $m_{H^\pm}=$ 150 GeV, $m_{N}=$ 100 GeV and BP2 as $m_{H^\pm}=$ 650 GeV, $m_{N}=$ 300 GeV respectively for each choice of $m_{Z^\prime}$. The cut flow for the SM backgrounds and the corresponding signals are given in Tab.~\ref{Signal:5leptons-1} where we estimate the combined significance for the signal of the corresponding BPs using
\begin{eqnarray}
\sigma = \sqrt{2\Big[(s+b)\log\Big(1+\frac{s}{b}\Big)-s\Big]},
\label{sig1}
\end{eqnarray}
where $s$ and $b$ stand for signal and background events at 3000 fb$^{-1}$ luminosity. Studying the signal and background processes over grids of 150 GeV $\leq m_{H^\pm}\leq 1.1$ TeV 
and 100 GeV $\leq m_N \leq 1$ TeV we estimate the significance contours for the combined 5 charged leptons in Fig.~\ref{fig:allowed gpp-1-1} at 3000 fb$^{-1}$ luminosity. We show the contours for 
$m_{Z^\prime}=3$ (5) TeV in the left (right) panel. The boundary of the light blue region shows 5$\sigma$ significance and the light blue region represents $m_{H^\pm}$ and $m_N$ which could be probed with a significance more than 5$\sigma$ but below 7$\sigma$ where the light cyan region represents $m_{H^\pm}$ and $m_N$ which could be probed with a significance more than 7$\sigma$ but below 10$\sigma$. The boundary of the blue shaded region represents $m_{H^\pm}$ and $m_N$ which could be probed with 10$\sigma$, however, the blue region represents $m_{H^\pm}$ and $m_N$ with more than 10$\sigma$. From Tab.~\ref{Signal:5leptons-1} we find that for $m_{Z^\prime}=3$ TeV BP1 and BP2 reside in the blue and cyan regions respectively. Similar results are obtained for $m_{Z^\prime}=5$ TeV. Above the black line $H^\pm\to \ell^\pm N$ is kinematically suppressed.
\begin{table}[h]
\centering
\begin{tabular}{|p{4.2cm}||p{2.3cm}|p{2.3cm}|p{2.3cm}|p{2.6cm}|p{3cm}|}
 \hline
 \multicolumn{6}{|c|}{Background cross sections for 5-muons at $\sqrt{s}=14$ TeV} \\
 \hline\hline
 Selection cuts & $ZZW^{\pm}$ [fb] & $t\tilde{t}ZZ$ [fb] & $t\tilde{t}WZ$ [fb] & $ZZW^+W^-$ [fb] & $ZW^\pm W^\pm W^\mp$ [fb]\\
 \hline
 Madgraph: & $3.650\times 10^{-3}$ & $1.392\times 10^{-4}$ & $1.153 \times 10^{-4}$ & $3.540\times 10^{-5}$ & $2.940\times 10^{-5}$ \\
 Basic cuts: & $7.033 \times 10^{-4}$ & $2.134\times 10^{-5}$ & $1.753\times 10^{-5}$ & $6.920 \times 10^{-6}$ & $6.851\times 10^{-6}$ \\
 Momentum cuts: & $3.597\times 10^{-4}$ & $1.321\times 10^{-5}$ & $1.083 \times 10^{-5}$ & $4.051\times 10^{-6}$ & $4.439\times 10^{-6}$ \\
 Lepton separation cuts: & $3.030\times 10^{-4}$ & $1.026\times 10^{-5}$ & $8.924 \times 10^{-6}$ & $2.996\times 10^{-6}$ & $3.393\times 10^{-6}$ \\
 \hline
  \hline
 \multicolumn{6}{|c|}{Background cross sections for 5-electrons at $\sqrt{s}=14$ TeV} \\
 \hline\hline
 Selection cuts & $ZZW^{\pm}$ [fb] & $t\tilde{t}ZZ$ [fb] & $t\tilde{t}WZ$ [fb] & $ZZW^+W^-$ [fb] & $ZW^\pm W^\pm W^\mp$ [fb] \\
 \hline
 Madgraph: & $3.650\times 10^{-3}$ & $1.392\times 10^{-4}$ & $1.153 \times 10^{-4}$ & $3.540\times 10^{-5}$ & $2.940\times 10^{-5}$ \\
 Basic cuts: & $2.399\times 10^{-4}$ & $6.675\times 10^{-6}$ & $5.628\times 10^{-6}$ & $2.424\times 10^{-6}$ & $2.574\times 10^{-6}$ \\
 Momentum cuts: & $1.233\times 10^{-4}$ & $4.178\times 10^{-6}$ & $3.565\times 10^{-6}$ & $1.420\times 10^{-6}$ & $1.742\times 10^{-6}$ \\
 Lepton separation cuts: & $1.028\times 10^{-4}$ & $3.159\times 10^{-6}$ & $2.922\times 10^{-6}$ & $1.017\times 10^{-6}$ & $1.319\times 10^{-6}$ \\
 \hline
\end{tabular}
\caption{Cross sections for the backgrounds of 5 charged leptons at $\sqrt{s} = 14$ TeV LHC after each  selection cuts.}
\label{BG:5muons}
\end{table}
\begin{table}[h]
\centering
\begin{tabular}{|p{4.cm}|p{1.5cm}|p{1.5cm}|p{1.5cm}|p{1.5cm}|p{1.5cm}|p{1.5cm}|}
\cline{1-7}
\multicolumn{5}{|c|}{Signal cross sections at $\sqrt{s} = 14$ TeV, $m_{Z^\prime}=3$ TeV}  & 
\multicolumn{2}{|c|}{Combined significance}\\  
\hline
\hline
\multicolumn{1}{|c|}{Selection cuts} & \multicolumn{2}{|c|}{5-muons [fb]} &  \multicolumn{2}{c|}{5-electrons [fb]}  & \multicolumn{1}{|c|}{BP1} & \multicolumn{1}{|c|}{BP2} \\
\cline{2-7}
& BP1 & BP2 &  BP1 & BP2 & & \\ \cline{1-5} 
Madgraph: & 2.699 & 0.0119 & 2.699 & 0.0119 & & \\ 
Basic cuts: & 0.223 & 0.00496 & 0.054 & 0.00180 & ~32.69 & ~9.18  \\ 
Momentum cuts: & 0.0515 & 0.00434 & 0.0132 & 0.00156 & & \\ 
Lepton separation cuts: & 0.0297 & 0.00369 & 0.00837 & 0.00133 & & \\ \cline{1-7}
\cline{1-7}
\multicolumn{5}{|c|}{Signal cross sections at $\sqrt{s} = 14$ TeV, $m_{Z^\prime}=5$ TeV}  & \multicolumn{2}{|c|}{Combined significance} \\ 
\hline
\hline
\multicolumn{1}{|c|}{Selection cuts} & \multicolumn{2}{|c|}{5-muons [fb]} &  \multicolumn{2}{c|}{5-electrons [fb]} & \multicolumn{1}{|c|}{BP1} & \multicolumn{1}{|c|}{BP2} \\
\cline{2-7}
& BP1 & BP2  & BP1 & BP2 & & \\ \cline{1-5}
Madgraph: & 2.662  & 0.0141 & 2.662 & 0.0141 & & \\ 
Basic cuts: & 0.206 & 0.00510 & 0.0471 & 0.00198 & ~28.63 & ~8.84 \\ 
Momentum cuts: & 0.0468 & 0.00437 & 0.00931 & 0.00162 & & \\ 
Lepton separation cuts: & 0.0258 & 0.00352 & 0.00479 & 0.00122 & &  \\ \cline{1-7}
\end{tabular}
\caption{Signal cross sections and significance for the 5 charged lepton events at $\sqrt{s} = 14$ TeV LHC after each selection cuts for the benchmark points BP1 and BP2 at a luminosity of 3000 fb$^{-1}$.}
\label{Signal:5leptons-1}
\end{table}
\subsubsection{Six charged lepton final state}
We discuss the 6 charged lepton final state from the heavy neutrino pair production induced by $H^\pm$ and followed by the dominant $N\to \ell W$ mode where $W$ bosons decay leptonically. In addition to that we consider the $N\to \nu Z$ mode from both the heavy neutrinos followed by leptonic decay of the $Z$ boson. Due to the lepton multiplicity, the 6 charged lepton final state will have small cross section. Therefore $N\to Z\nu$ mode is added to boost the signal because BR$(N\to Z\nu)$ is roughly $25\%$ for $m_{N} \geq 141$ GeV which can be found in Fig.~\ref{BR-N}. 

We generate irreducible backgrounds for 6 charged leptons which could be fully muons or fully electrons. Here we have the events with 6 charged leptons in association with missing momentum where jets may appear from the initial state radiations. The irreducible backgrounds are: (i) $ZZZ$ where three $Z$ bosons decay leptonically, (ii) $t\tilde{t}ZZ$ where $Z$ decays leptonically, $t\tilde{t}$ produce $W^\pm$ with b-jets and the $W$ bosons decay leptonically, (iii) $t\tilde{t}WWZ$: the $W$ and $Z$ decay leptonically and $t\tilde{t}$ produce $W^\pm$ in association with $b-$jets and finally the $W^\pm$ bosons decay leptonically, (iv) $ZZW^+W^-$: the $Z$ and $W$ bosons decay leptonically. In this analysis additionally we consider a reducible background coming from $ZZ+$jets. In this case 4 charged leptons are coming from the $Z$ bosons and the remaining ones come from the misidentified jets. Identification of two mistagged leptons from jets is negligibly small. Hence without $Z$-veto, the $ZZ$+jets background will be very small for 6 charged lepton final state. Hence we ignore this background. To study the signal and backgrounds at the 14 TeV LHC we impose the following selection cuts:
\begin{enumerate}
\item {\it Basic Cuts:} We select events with 6 leptons (fully muons/electrons) in the final state with minimum transverse momentum $p_T(\ell) >$ 10 GeV and $|\eta_\ell|<$ 2.4. Minimum transverse momentum for the accompanying radiated jets to be, $p_T(j)> $ 20 GeV and $|\eta_j|<$ 2.5.
\item {\it Momentum cuts:} To reduce the backgrounds further we impose a hard cut on the transverse momentum of the leptons. We impose $p_T(\ell_{1,2}) > 30$ GeV for the leading leptons and $p_T(\ell_{3,4,5, 6}) > 20$ GeV for the trailing leptons respectively where leptons are ordered according to the transverse momentum. Along with that we impose a missing momentum cut of $p_T^{\rm miss} > 30$ GeV to reduce events with jets as fake leptons further.
\item {\it Lepton separation cuts:} We impose the lepton-lepton separation in the $\eta-\phi$ plane as $\Delta R_{\ell \ell} > 0.4$ and if there are any accompanied jets, then $\Delta R_{\ell j} > 0.4$ and $\Delta R_{j j} > 0.4$ have been considered where $\Delta R = \sqrt{\Delta \eta^2 + \Delta \phi^2}$.   
\end{enumerate}
Additional cuts like $b$-veto and $Z$-veto reduce the top quark and $Z$ boson mediated backgrounds significantly, however since the backgrounds are already negligible due to the low cross sections, the above mentioned cuts are sufficient enough. The background cross sections based on the above cuts are mentioned in Tab.~\ref{BG:6muons} for the case of 5 fully muons/electrons final states. For the signal we choose the benchmark parameters as $m_{Z^\prime}=3$ TeV and 5 TeV, $m_{H^\pm}=$ 150 GeV and 650 GeV with $m_N=100$ GeV and 300 GeV respectively. We recognize Benchmark Points (BPs): BP1 as $m_{H^\pm}=$ 150 GeV, $m_{N}=$ 100 GeV and BP2 as $m_{H^\pm}=$ 650 GeV, $m_{N}=$ 300 GeV respectively for each choice of $m_{Z^\prime}$. The cut flow for the SM backgrounds and the corresponding signals are given in Tab.~\ref{Signal:6leptons-1} where we estimate the significance of the signal using Eq.~\ref{sig1} at 3000 fb$^{-1}$ luminosity.

\begin{figure}[h]
\includegraphics[width=89mm]{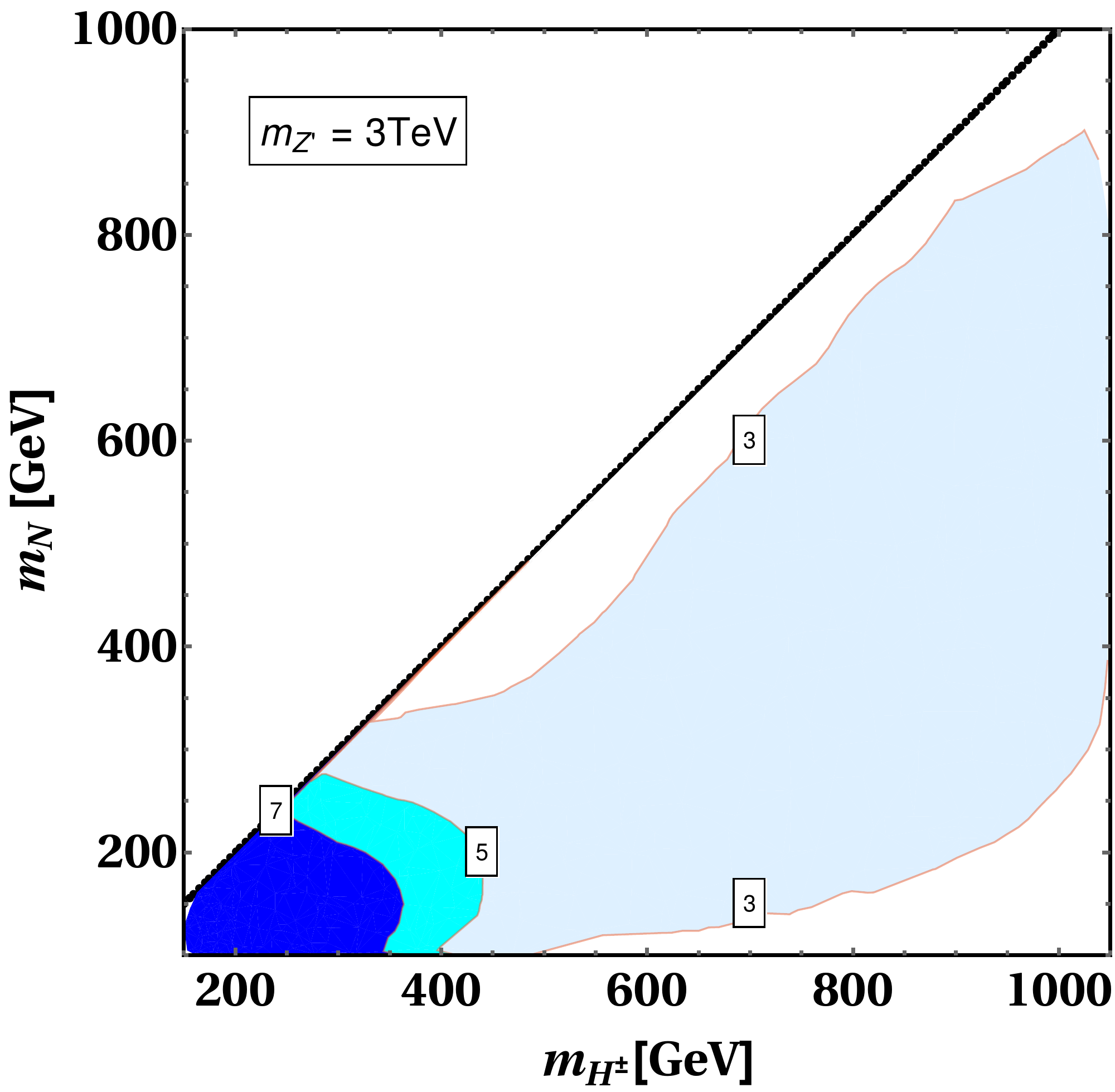}
\includegraphics[width=89mm]{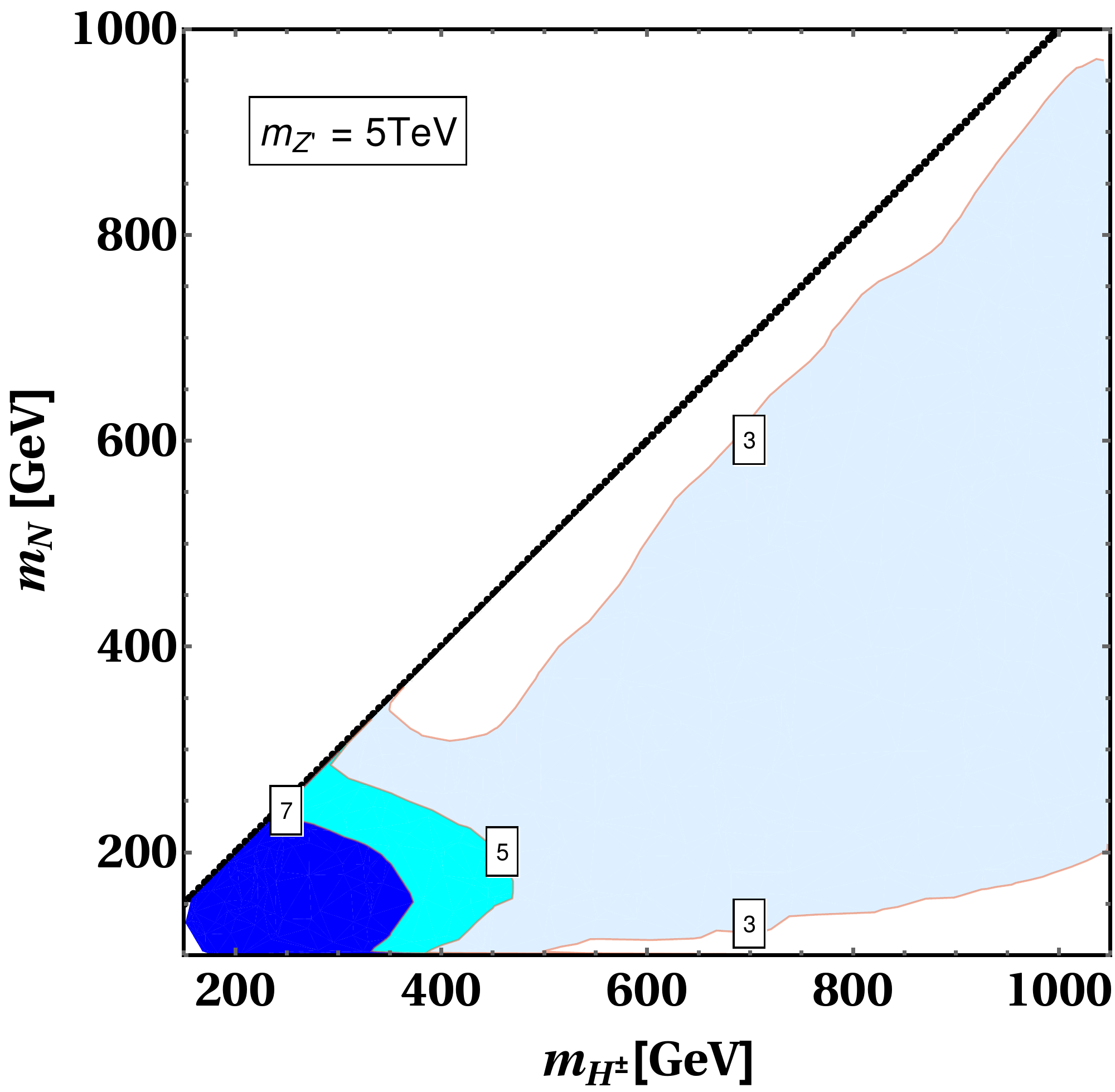}
\caption{Different significance contours for the 6 charged lepton final state with 3000 fb$^{-1}$ luminosity. The black straight line represents $m_{H^{\pm}}=m_N$, below which the on-shell decay of $H^\pm$ occurs.}
\label{fig:allowed gpp-1-2}
\end{figure}

Studying the signal and background processes over grids of 150 GeV $\leq m_{H^\pm}\leq 1.1$ TeV and 100 GeV $\leq m_N \leq 1$ TeV we estimate the significance contours for the combined 6 charged leptons in Fig.~\ref{fig:allowed gpp-1-2} at 3000 fb$^{-1}$ luminosity. We show the contours for $m_{Z^\prime}=3$ (5) TeV in the left (right) panel. The boundary of the light blue region shows 3$\sigma$ significance and the light blue region represents $m_{H^\pm}$ and $m_N$ which could be probed with a significance more than 3$\sigma$ but below 5$\sigma$ where the light cyan region represents $m_{H^\pm}$ and $m_N$ which could be probed with a significance more than 5$\sigma$ but below 7$\sigma$. The boundary of the blue shaded region represents $m_{H^\pm}$ and $m_N$ which could be probed with 7$\sigma$, however, the blue region represents $m_{H^\pm}$ and $m_N$ with more than 7$\sigma$. From Tab.~\ref{Signal:6leptons-1} we find that for $m_{Z^\prime}=3$ TeV BP1 and BP2 reside in the blue and cyan regions respectively. Similar results are obtained for $m_{Z^\prime}=5$ TeV. Above the black line $H^\pm\to \ell^\pm N$ is kinematically suppressed.
\begin{table}[h]
\centering
\begin{tabular}{|p{4.2cm}||p{2.3cm}|p{2.3cm}|p{2.3cm}|p{2.6cm}|}
 \hline
 \multicolumn{5}{|c|}{Background cross sections for 6-muons at $\sqrt{s}=14$ TeV} \\
 \hline\hline
 Selection cuts & $ZZZ$ [fb] & $t\tilde{t}ZZ$ [fb] & $t\tilde{t}WWZ$ [fb] & $ZZWW$ [fb] \\
 \hline
 Madgraph: & $3.949 \times 10^{-4}$ & $1.897\times 10^{-5}$ & $3.815\times 10^{-7}$ & $5.830\times 10^{-6}$  \\
 Basic cuts: & $6.553\times 10^{-5}$ & $2.614\times 10^{-6}$ & $4.586\times 10^{-8}$ & $1.164\times 10^{-6}$ \\
 Momentum cuts: & $3.451\times 10^{-6}$ & $1.594\times 10^{-6}$ & $2.703\times 10^{-8}$ & $6.928\times 10^{-7}$  \\
 Lepton separation cuts: & $2.207\times 10^{-6}$ & $1.179\times 10^{-6}$ & $1.810\times 10^{-8}$ & $4.498\times 10^{-7}$ \\
 \hline
  \multicolumn{5}{|c|}{Background cross sections for 6-electrons at $\sqrt{s}=14$ TeV} \\
 \hline\hline
 Selection cuts & $ZZZ$ [fb] & $t\tilde{t}ZZ$ [fb] & $t\tilde{t}WWZ$ [fb] & $ZZWW$ [fb] \\
 \hline
 Madgraph: & $3.949\times 10^{-4}$ & $1.897\times 10^{-5}$ & $3.815 \times 10^{-7}$ & $5.830\times 10^{-6}$  \\
 Basic cuts: & $1.906 \times 10^{-5}$ & $7.051 \times 10^{-7}$ & $1.215 \times 10^{-8}$ & $3.704\times 10^{-7}$ \\
 Momentum cuts: & $6.713 \times 10^{-7}$ & $4.029\times 10^{-7}$ & $7.692\times 10^{-9}$ & $2.292\times 10^{-7}$  \\
 Lepton separation cuts: & $4.462 \times 10^{-7}$ & $2.887\times 10^{-7}$ & $5.188\times 10^{-9}$ & $1.463 \times 10^{-7}$ \\
 \hline
\end{tabular} 
 \caption{Cross sections for the backgrounds of 6 charged leptons at $\sqrt{s} = 14$ TeV LHC after each selection cuts.}
\label{BG:6muons}
\end{table}
\begin{table}[h]
\centering
\begin{tabular}{|p{4.cm}|p{1.5cm}|p{1.5cm}|p{1.5cm}|p{1.5cm}|p{1.5cm}|p{1.5cm}|}
\hline
\multicolumn{5}{|c|}{Signal cross sections at $\sqrt{s} = 14$ TeV, $m_{Z^\prime}=3$ TeV}  & 
\multicolumn{2}{|c|}{Combined significance}\\  
\hline 
\hline
\multicolumn{1}{|c|}{Selection cuts} & \multicolumn{2}{|c|}{6-muons [fb]} &  \multicolumn{2}{c|}{6-electrons [fb]}  & \multicolumn{1}{|c|}{BP1} & \multicolumn{1}{|c|}{BP2} \\
\cline{2-7}
& BP1 & BP2 &  BP1 & BP2 & & \\ \cline{1-5} 
Madgraph: & 0.430 & 0.00220 & 0.430 & 0.00220 & & \\ 
Basic cuts: & 0.0698 & 0.000851 & 0.0191 & 0.000269 & ~21.68 & ~4.54  \\ 
Momentum cuts: & 0.0253 & 0.000741 & 0.00731 & 0.000238 & & \\ 
Lepton separation cuts: & 0.00851 & 0.000569 & 0.00232 & 0.000179 & & \\ \hline
\hline
\multicolumn{5}{|c|}{Signal cross sections at $\sqrt{s} = 14$ TeV, $m_{Z^\prime}=5$ TeV}  & \multicolumn{2}{|c|}{Combined significance} \\ 
\hline
\hline
\multicolumn{1}{|c|}{Selection cuts} & \multicolumn{2}{|c|}{6-muons [fb]} &  \multicolumn{2}{c|}{6-electrons [fb]} & \multicolumn{1}{|c|}{BP1} & \multicolumn{1}{|c|}{BP2} \\
\cline{2-7}
& BP1 & BP2  & BP1 & BP2 & & \\ \cline{1-5}
Madgraph: & 0.424 & 0.0026 & 0.424 & 0.0026 &  & \\ 
Basic cuts: & 0.0722 & 0.000949 & 0.0198 & 0.000321 & ~19.86 &~4.87 \\ 
Momentum cuts: & 0.0256 & 0.000847 & 0.00716 & 0.000292 & & \\ 
Lepton separation cuts: & 0.00763 & 0.000626 & 0.00165 & 0.000214 & &  \\ \cline{1-7}
\end{tabular}
\caption{Signal cross sections and significance for the 6 charged lepton events at $\sqrt{s} = 14$ TeV LHC after each selection cuts for the benchmark points BP1 and BP2 at a luminosity of 3000 fb$^{-1}$.}
\label{Signal:6leptons-1}
\end{table}
\section{Conclusion}
\label{conc}
In this paper we consider a scenario where SM is extended by a general U$(1)_X$ gauge group. To cancel the gauge and mixed gauge gravity anomalies we introduce three generations of RHNs with nonuniversal charge assignments under U$(1)_X$ gauge group. Due to the U$(1)_X$ gauge symmetry the RHNs interact with the extended Higgs sector of the SM giving rise to the seesaw mechanism to generate the light neutrino mass after the U$(1)_X$ symmetry is broken. In this scenario a neutral gauge boson $Z^\prime$ is evolved where the left and right handed fermions, being charged differently under general U$(1)_X$, interact with the $Z^\prime$ manifesting the chiral nature of the model. Thanks to the general U$(1)_X$ charge assignment which is responsible for the dominant decay mode of $Z^\prime$ into a pair of charged Higgs over its dilepton decay mode. Hence for that U$(1)_X$ charge assignment, the experimental bounds on the gauge coupling for different $M_{Z^\prime}$ obtained from the dilepton searches at the LHC and LEP-II will be comparatively loose. It will help to enhance charged Higgs pair production cross section. Meanwhile, in the limit of low $\tan\beta$ we find that the second Higgs becomes neutrinophilic and its charged multiplet completely decays into a charged lepton and RHN. Producing the charged Higgs in pair and studying its decay into RHNs we obtain rare multilepton final states. Here we consider the decay of the RHNs into the dominant mode $\ell W$.  Following the hadronic and/or leptonic decay of the $W$ boson we obtain five or six charged leptons in the final state. This leptonic multiplicity helps us to probe such rare multilepton final states from the RHNs. Comparing the signals and irreducible backgrounds for different regions in $m_{H^\pm}-m_N$ plane we obtain significantly high discovery potential for these final states at the 14 TeV LHC with 3000 fb$^{-1}$ luminosity. The six charged lepton final state is weaker than the five charged lepton final state beacsue leptonic decay of the $W$ boson reduces cross section. We comment that in this article we have studied purely electron and muon signals. However, it is possible to study electron and muon both in the final state which may come from the charged Higgs, heavy neutrino and the leptonic decay of the $W$ boson. This will definitely increase the significance of the signal over the backgrounds, however, we already have a clean final state due to lepton multiplicity. Hence due to simplicity we did not consider the mixed signal.  In future several interesting aspect of including collimated objects, displaced scenarios could be tested from the charged Higgs induced RHN productions in the context of hadron colliders. The reconstruction of the $Z^\prime$ boson will be another interesting aspect at the high energy colliders where all the final states are visible involving leptons and jets.
\begin{acknowledgments}
PS would like to thank Takaaki Nomura for useful discussions. The work of SK is supported by JSPS KAKENHI 20H00160 and 21F21324.
The work of PS is supported by an appointment to the JRG Program at the APCTP through the Science and Technology Promotion Fund and Lottery Fund of the Korean Government. This is also supported by the Korean Local Governments - Gyeongsangbuk-do Province and Pohang City.
\end{acknowledgments}
\bibliography{bibliography}
\end{document}